\documentclass[12pt]{iopart}

\usepackage{iopams}
\usepackage{amssymb,amsfonts}
\usepackage{graphicx}
\graphicspath{{Figures/}}
\usepackage{psfrag}
\def\contentsname{}
\usepackage{bm}

\begin{document}

\bibliographystyle{unsrt}

\title[Possible Role of Interference and Sink Effects \dots]{
Possible Role of Interference and Sink Effects in  Nonphotochemical Quenching in Photosynthetic Complexes }

\author{Gennady P.  Berman}

\address{Theoretical Division, T-4, MS-B213, Los Alamos National Laboratory, and the New Mexican Consortium, 100 Entrada Dr., Los Alamos, NM 87544, USA}
 \ead{gpb@lanl.gov}

\author{Alexander I. Nesterov }

\address{Departamento de F{\'\i}sica, CUCEI, Universidad de Guadalajara,
Av. Revoluci\'on 1500, Guadalajara, CP 44420, Jalisco, M\'exico}
\ead{nesterov@cencar.udg.mx}

\author{Shmuel Gurvitz}
 
 \address{Department of Particle Physics and Astrophysics, Weizmann Institute,
76100, Rehovot, Israel}
\ead{shmuel.gurvitz@weizmann.ac.il}

\author{ Richard T.~Sayre}

\address{ Los Alamos National Laboratory and the New Mexican Consortium, 100 Entrada Dr., Los Alamos, NM 87544, USA}
 \ead{rsayre@newmexicoconsortium.org}
   
\submitto{\NJP}

\begin{abstract}
We describe a simple and consistent quantum mathematical model that simulates the  possible role of quantum interference and sink effects in the nonphotochemical quenching (NPQ) in  light-harvesting complexes (LHCs). Our model consists of a network of five interconnected sites (excitonic states)  responsible for the NPQ mechanism: (i) Two excited states of chlorophyll molecules, $ChlA^*$ and $ChlB^*$, forming an LHC dimer, which is initially populated; (ii) A ``damaging" site which is responsible for production of singlet oxygen and other destructive outcomes; (iii)  The $(ChlA-Zea)^*$ heterodimer excited state (Zea indicates zeaxanthin); and (iv) The charge transfer state  of this heterodimer,  $(ChlA^{-}-Zea^{+})^*$. In our model, both damaging and charge transfer states are described by discrete electron energy levels attached to their sinks, that mimic the continuum part of electron energy spectrum, as at these sites the electron participates in quasi-irreversible chemical reactions. All five excitonic sites interact with the protein environment that is modeled using a stochastic approach. As an example, we apply our model  to demonstrate possible contributions of quantum interference and sink effects in the NPQ mechanism in the CP29 minor LHC. Our numerical results on the quantum dynamics of the reduced density matrix,  demonstrate a possible way to significantly suppress, under some conditions, the damaging channel using quantum interference effects and sinks. The results demonstrate the possible role of interference and sink effects for modeling, engineering, and optimizing the performance of the NPQ processes in both natural and artificial light-harvesting complexes.
\end{abstract}

\maketitle

\tableofcontents


\maketitle

\section{Introduction}
\label{intro}
The first step in photosynthesis is the capture of photons, which occurs predominantly in the light-harvesting complexes (LHCs). The LHCs of the photosystems I and II (PSI and PSII) consist of hundreds of light-sensitive, protein bound chlorophyll molecules.  A quantum of sunlight, which is absorbed by a chlorophyll molecule,  has rather high energy, on the order of $1.5-2.5eV$. Nature has developed a mechanism to transfer this locally excited electron energy (exciton) very efficiently ($\sim 95\%$) from the peripheral LHC to the reaction center (RC), where  charge separation and complex chemical reactions take place. Because a network of the exciton transfer (ET) through the  LHC to the RC includes many light-sensitive molecules and is rather complicated and quasi-random,   the dynamics of the ET must be very rapid in order for the exciton not to be lost to fluorescence, thermal fluctuations of the protein environment, and other electromagnetic processes. Indeed, the dynamics of the ET is rather fast, from one to tens of picoseconds. It was recently demonstrated, that quantum coherent effects survive (and significantly enhance the efficiency) in LHCs and RCs during the primary processes of ET for times of order of hundreds of femtoseconds, even at room temperature \cite{Vos,Hu,BS,Lee,Engel,Ras,Pan,Col,Wes,Wong,Ish,Ch,Sch,Str,Ferretti}. (See also references therein.)

The structure of the  LHC PSII (supercomplex) is very complicated, and includes more than 300 light-sensitive chlorophyll ($Chla$ and $Chlb$)  molecules, which contain excitonic sites (similar to connected quantum dots in solids) connected in a very complicated network. (See, for example,  \cite{Bennett}, and references therein.)  These excitonic sites are embedded in a protein environment that operates as (i) the thermal bath and noise generator for excitonic sites of light-sensitive $Chls$ and (ii) an external media that changes in time its geometrical configuration (conformations) into different operational modes. In particular, it is well-known that if even one amino acid in the protein sequence is modified, this can potentially change the protein conformation or alter the energy levels of $Chls$ and their ability to function. These conformational  modifications result in changes of locations and energy levels of the light-sensitive sites including interacting carotenoids and, consequently, the light-harvesting properties of the whole system \cite{Harrop}.

In order to better characterize the ET properties inside the peripheral light-harvesting supercomplex, it is conventionally divided into (i) the most peripheral ``major" LHCs \cite{Bennett,Schlau} which have a  trimeric structure, and (ii) minor complexes more closely associated with the PSII RC which include  CP24, CP26, CP29, CP43, and CP47, located between the RC and the peripheral LHCII complex \cite{Bennett}.  All these LHCII and minor complexes include a network of chlorophylls and carotenoids that  absorb the solar quanta, and, as it was mentioned above,  transfer the excitons from the antenna to the RC.

Note, that the sun energy is so diluted, that when modeling the exciton dynamics, under  normal sunlight intensity, it is often assumed that only the lowest-lying singlet excited states of chlorophylls are required \cite{Bennett}. At the same time, when the solar flux levels exceed the capacity of downstream electron transfer processes, the RC becomes saturated, and several ``damaging" channels are opened. These damaging channels result in unwanted photo-damaging effects, that can destroy the whole organism. This process is called photoinhibition \cite{Mull,Li}.

Photosynthetic organisms have evolved many mechanisms which protect them from damaging processes associated with high sunlight intensity \cite{Li1}. One of these mechanisms, is ``nonphotochemical quenching" (NPQ). NPQ is induced by the low pH of the thylakoid lumen and involves multiple steps including the pH dependent de-epoxidation of violoxanthin to zeaxanthin to dissipate excess excited states as heat.  Many scientific teams have been involved for years in theoretical and experimental analysis, and in mathematical modeling of these NPQ mechanisms  \cite{Dur,Mel,Ander,Mul,And2,Dreu,Dreuw1,Holt,Dreu2,Holt2,Cheng,Ahn,Gil,Ameron,Ma,Val,Bia,Gor,Vaz,Zak,Duffy1,Duffy2,Rub,Zak2,Lamb,Hol,Zak3,Croce}. (See also references therein.) But because the LHC network is very complicated and because NPQ mechanisms involve many complex physiological aspects of the photosynthetic organisms, still neither the detailed location of excitonic sites inside the antenna nor  the detailed mechanisms  responsible for quenching processes have been established. For example, models describing the NPQ mechanisms vary, from relatively simple, which involve three \cite{Holt2} or four \cite{Cheng} excitonic and charge transfer states, to very complicated, which involve 26 nonlinear ordinary differential equations with 78 fitting parameters \cite{Zak}.

At the same time, some components of the NPQ are generally accepted. For example, it is generally believed that NPQ is regulated by protonation processes and, correspondingly, by the magnitude of the trans-membrane pH gradient across the thylakoid membrane. (See, for example, \cite{Dreu2,Zak,Zak3}, and references therein.) It is also believed by many that the NPQ requires the presence of the antenna-associated PsbS (plants) or LhcSR (green algae) and other proteins which regulate the NPQ process by as of yet unknown mechanisms  \cite{Li1,Ander,Mul,And2,Dreu,Dreu2,Cheng,Ahn,Bia,Gor,Zak,Duffy1,Rub,Zak2,Lamb,Hol,Zak3,Peers,Alb,Bai,Niy,Bonen}. 

Overall, however, it would be useful to have a simple and consistent quantum-mechanical mathematical model which (i) is {\it linear} for the reduced density matrix elements (as required by quantum mechanics) and (ii) allows one to analyze, describe and better understand some important fragments of these very complicated and multi-functional  NPQ mechanisms.

In this paper, we introduce a simple quantum mathematical model which allows us to simulate a possible role of quantum interference (coherent) effects and  quantum effects due to the interaction of discrete energy levels (associated with light-sensitive sites) with the continuum parts of the electron energy spectrum (sinks), in the NPQ mode.  
For purposes of illustration of our approach, our  model is based on the NPQ analysis used in \cite{Dreuw1,Dreu2,Cheng,Ahn}, in which the formation of the carotenoid zeaxanthin (Zea) and a creation of the $(Chl-Zea)^*$ heterodimer and charge transfer state (CTS), $(ChlA^{-}-Zea^{+})^*$, were discussed in the CP29 and others minor antenna complexes, in the NPQ phase. (For creation of the $(Chl-Zea)^*$ heterodimer see, for example, \cite{Gil,Ameron,Ma}.) In particular, in \cite{Ahn}, one possible  role of the charge-transfer quenching (CTQ) was considered in three minor antenna complexes, CP29, CP26, and CP24. The authors of \cite{Ahn} argued that the CTQ in CP29 involves (i) a delocalized state of the {\it strongly} coupled dimer of the two chlorophylls, $ChlA_5$ and $ChlB_5$, (ii) a heterodimer which includes the $Zea$ molecule, and (iii) the CTS of this heterodimer.

Our model includes five excited discrete excitonic and charge transfer states which could be responsible for the NPQ mechanism: (i) Two excited states of chlorophylls,  $ChlA^*$ and $ChlB^*$, which can be initially populated either in the excited state of a single  $ChlA^*$ or $ChlB^*$ (when the dimer is weakly coupled), or in the excited eigenstate(s) of the dimer, $(ChlA-ChlB)^*$, when the dimer is strongly coupled; (ii) A damaging excited electron state, $D^*$, which is responsible for production of damaging outcomes such as very reactive singlet oxygen, and others; (iii)  The  $(ChlA-Zea)^*$ heterodimer excited state; and (iv) The CTS of this heterodimer, $(ChlA^{-}-Zea^{+})^*$.

In our model, the two sites, $D^*$ and $(ChlA^{-}-Zea^{+})^*$, are described by  discrete electron energy levels interacting with their continuum spectra (sinks). This approach is in many aspects relevant to modeling of unstable quantum systems, such as heavy nuclei, solid-state nano-systems, and others \cite{Zel1,Zel2,Ber1,Ber2,Ber3,Ber4}.  In this case,  the initial system (reactant) undergoes an irreversible (or quasi-irreversible, with possible recurrences) decay through some channels (reactions) into the final state(s) (product(s)).  We associate one sink with the damaging state, $|0\rangle$, and another sink with the CTS, $|4\rangle$.  These two sinks play, in some respect,  a similar role  as two potential quenching channels, but with particular functions.

We assume, that when the light intensity is high enough, the PSII RC is closed by the mechanism(s) similar to the Coulomb blockade regime in a quantum dot, in solids. Namely, if one electron populates already the small quantum dot (RC), the second electron cannot occupy the same energy level in the dot, but could occupy the next excited level in the dot, which usually requires large additional energy, if the dot is small. But because at high light intensity, the number of photons that correspond to the first excited chlorophyll state increases, the next exciton may appear. In this case, when this ``extra" exciton populates the LHC chlorophyll dimer, its electron cannot be transfered to the PSII RC. Instead, it can be transfered to the damaging state, $|0\rangle$, or to participate in the NPQ, or to be involved in many different processes. So, in our model, we do not use directly the highly excited states of the chlorophyll molecules. But they can  easily be included in our model.

The electron of the initially populated exciton, in the NPQ regime, exists in different ``forms": (i) It can belong to the exciton which travels through the network of the LHC, and only slightly reducing (or accumulating from the environment) its energy; (ii) It can be absorbed by the damaging sink, $|S_0\rangle$, and participate in creation of damaging product states; or (iii) Can be absorbed by the sink, $|S_4\rangle$, of the CTS, and participate in non-harmful chemical reactions. This last option is called a CTQ. Certainly, according to quantum mechanics, this electron can also exist in a superposition of all these and other states of our model. Each sink in our model is characterized by two parameters: the rate, $\Gamma_{0}$ ($\Gamma_{4}$), of transfer into the sink, $|S_{0}\rangle$ ($|S_{4}\rangle$), from the corresponding attached discrete state, $|0\rangle$ ($|4\rangle$), and the efficiency (cumulative time-dependent probability), $\eta_{0}(t)$ ($\eta_{4}(t)$), of being absorbed by this sink.  Note, that both $\Gamma_0$ and $\Gamma_4$, characterize only the rates of {\it destruction} of the exciton in the LHC, and do not describe the following processes with the occurred electron in the damaging and the charge transfer reactions. In this sense, our model describes only the primary NPQ processes related to the ET, and does not describe the processes which occur in both sinks. The later occur on relatively large time-scales, and require a detailed knowledge of the structures of the sinks, and additional methods for their analysis.

All five excitonic sites interact with their protein environment, which we model by the stochastic process of stationary telegraph noise. Note, that there are many different approaches for modeling the influence of the protein environment on the excitonic and electron sites in bio-systems \cite{Pan,Wes,Ish,Ch,Schlau,Marcus1,Xu,Deway,Ben,Pudlak1,Pudlak2,Pudlak3,Pudlak4,Sk,Das,Nes1,Nes2,Ber5,Ish2,Lloyd1,Moix,Abramavicius,Nest4}. Some of them use a thermal bath (for example, as a set of harmonic oscillators in the initial Gibbs distribution, at a given temperature); some approaches use the external sources of noise; and some approaches are based on hybrid schemes \cite{Moix,Abramavicius}.  From our viewpoint, in the case of  bio-systems in a {\it non-functional} mode, it is reasonable to use a thermal bath as the source of fluctuations which the electron sites experience from the protein environment. But when bio-systems are operating in their functional modes, the question of the proper modeling of the action of protein environment on embedded network of excitonic and electron sites is more complicated. It is most likely that some environmental modes act on the excitonic and electron sites as  thermal bath modes, some as external noise (generators), and some even as regular forces. In particular, this can be the case for very fast primary ET dynamics in which many modes of the environment do not have time to approach their equilibrium states. (See also discussions on this issue in \cite{Pudlak1,Renger}.) We have chosen the model of the random telegraph noise mainly for two reasons: (i) We believe that the influence of the interference and sink effects on the NPQ process, which we consider, can be reasonably described by a stochastic environment.  
(ii) The set of linear ordinary differential equations for the ET, which we derive, is exact and closed. This allows us to derive ``exact" numerical solutions, and to avoid dealing with the uncertainties caused by different approximations used. This is especially important when the interaction constants between the excitonic sub-system and the environment are sufficiently large, as often happens in bio-systems at room temperature \cite{Marcus1,Xu}, and it becomes impossible to derive an exact and closed system of equations for the excitonic (and electron) reduced density matrix. 

In the NPQ regime, when structural changes in the CP29 complex occur (in particular, due to protonation, PsbS protein involvement, and other processes 
\cite{Ander,Mul,And2,Dreu,Dreuw1,Dreu2,Cheng,Ahn,Bia,Gor,Zak,Rub,Zak2,Lamb,Hol,Zak3}), the excited state(s) of the  LHC chlorophyll dimer, $(ChlA-ChlB)^*$, is realized and is initially populated. In this case, we  numerically find the conditions required to suppress the damaging channel,  for both limits of weakly and strongly coupled $(ChlA-ChlB)^*$ LHC dimers. In particular, for the case of a strongly coupled dimer, we demonstrate the suppression of the damaging channel caused by the simultaneous action of quantum interference effects and sinks. 

In order to demonstrate our results, we explicitly calculate all averaged over noise density matrix elements for the discrete states as well as the ``efficiencies" of both the damaging channel, $\eta_0(t)$, and the charge transfer channel, $\eta_4(t)$. For concreteness, we apply our results to describe a possible contribution of quantum interference and sink effects in the NPQ mechanism in the CP29 minor complex. At the same time, we should note that we do not intend here to fit our numerical results with the experimental results on the NPQ process in the CP29 complex. There are two main reasons for this. First, we do not think that this fitting is reasonable at present time, because  many details are currently unknown, such as  the location of the responsible excitonic sites and states, characteristics of the LHC dimer and heterodimer in the NPQ mode, the values of rates between sites and to the sinks, and other important parameters. Second, by varying many fitting parameters, (such as energy levels, values of interaction matrix elements, rates to the sinks, interaction constants between sites and noise, decay rate of noise correlations), one can match the particular results of numerical simulations with the particular experimental results. Instead, our main intention here is to discuss a possible role of quantum interference and sink effects in the NPQ mechanism. One advantage of our model is that it can be easily generalized including as many as needed excitonic states, CTSs and damaging channels, much more complex networks of interconnected excitonic sites, and a proper combination of thermal environment and external noise, in order to model the protein environment in the NPQ mode.\\
Our main results include:\\ \ \\
1. An exact and closed system of linear ordinary differential equations for the averaged over noise excitonic density matrix, which describes the influence of quantum interference effects, sinks, and noise on the NPQ process in the LHCs.\\ 
2.	Numerical experiments for a variety of different parameters and initial conditions of our model. We demonstrated the regimes of suppression of the damaging channel for both weakly and strongly coupled $(ChlA-ChlB)^*$ LHC dimer.\\ 
3. A demonstration of how, for the strongly coupled $(ChlA-ChlB)^*$ LHC dimer,  interference effects play a significant role in blocking the damaging channel.\\
4.	A quantitative parameter which characterizes the magnitude of blocking the damaging channel. This parameter is the probability of finding the electron in the sink of the damaging channel (efficiency).  If this parameter equals unity, the damaging channel is completely open. If this parameter is zero, the blocking of the damaging channel is 100\%.\\
5.	 A demonstration of the significant contribution of sinks in the performance of the NPQ mechanism, and an effectiveness of utilizing simultaneously both  interference and sinks effects for strongly coupled $(ChlA-ChlB)^*$ LHC dimer, in the NPQ mode.\\
6. A demonstration, for a weakly coupled LHC dimer,  of the  robustness of the NPQ regime in relation to the initial population of the LHC dimer.\\  
7.	A demonstration of the presence of collective quantum coherent effects, in the NPQ process.

The paper is organized as follows. In Sec. II, we describe our model and introduce its main characteristics. In Sec. III, we present the mathematical formulation of our model. In Sec. IV, we describe the approach based on noise-assisted ET. In Sec. V, we present the characteristic properties of the LHC dimer. In order to better characterize the interference effects in the NPQ regime, we consider in Sec. VI,  a simplified three-level system. The results of numerical simulations for the total system are presented in Sec. VII.
 In the Conclusion, we summarize our results. Appendices, A and B, include the derivation of the main equations.

\section{Description of the model}
\label{sec:1}

Our model includes five discrete excitonic energy states (see Fig. \ref{S2}):
 $|n\rangle$, $n=0,...,4$, and two sinks, $|S_0\rangle$  and  $|S_4\rangle$. The sink,  $|S_0\rangle$, interacts with the ``damaging state",  $|0\rangle$.  The sink, $|S_4\rangle$, interacts with the CTS, $|4\rangle$. All discrete states are interconnected through the matrix elements, $V_{mn}$, in the Hamiltonian (\ref{Ham}). All discrete energy levels interact with the external protein noisy environment.
 \begin{figure}[tbh]
 \begin{center}\label{S2}
 \scalebox{0.375}{\includegraphics{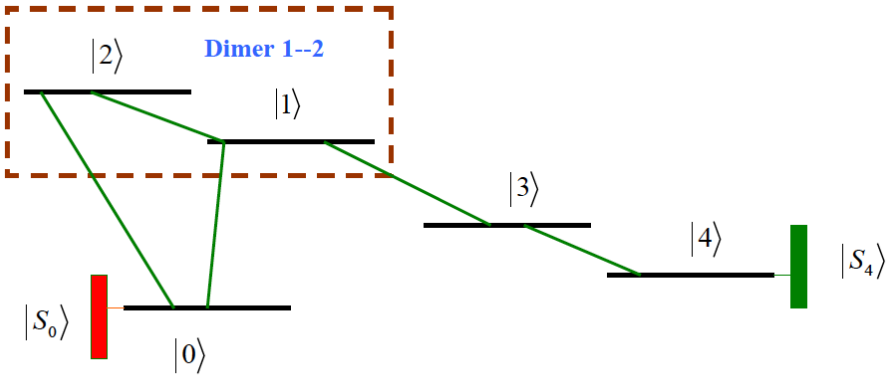}}
 \end{center}
 \caption{Schematic of our  model consisting of five discrete  states, $|n\rangle$, $(n=0,...,4)$, and two independent sink reservoirs, $|S_0 \rangle$ (connected to the damaging state), and  $|S_4 \rangle$ (connected to the charge transfer state). The rectangle indicates the initially populated dimer, $(ChlA-ChlB)^*$. 
 \label{S2}}
  \end{figure}

Below, for concreteness, we will use the notations and the basic components responsible for the NPQ process similar to those used in \cite{Cheng,Ahn}.  But the values of parameters will be varied in order to demonstrate the different dynamical regimes in our model, related to the performance of the NPQ mechanism. We use the following notation. The discrete  electron state, $|1\rangle\equiv |ChlA_5^*\rangle$, is the excited electron state of  chlorophyll, $A_5$.  The discrete  electron state, $|2\rangle\equiv |ChlB_5^*\rangle$, is the excited electron state of chlorophyll, $B_5$.  So, in our model only two chlorophylls, $A_5$ and $B_5$, each with an excited electron state, participate in the NPQ process, as in \cite{Ahn}. The discrete state,  $|3\rangle\equiv |(ChlA_5-Zea)^*\rangle$, is the heterodimer excited state.  The discrete state,  $|4\rangle\equiv |(ChlA_5^{-}-Zea^{+})^*\rangle$, is the CTS of the heterodimer. The state, $|0\rangle\equiv |damage\rangle$, is a discrete part of the damaging state (channel). 

Fig. \ref{S2} demonstrates the NPQ regime in which  (i) the weakly or strongly coupled dimer (shown in the rectangle) is realized based on two site states, $|1\rangle$ and $|2\rangle$, (ii) initially one of two eigenstates of the dimer is populated. (For a weakly coupled dimer, this corresponds approximately to the initial population, either $ChlA^*$ ($|1\rangle$) or $ChlB^*$ ($|2\rangle$).)

The sink, $|S_0\rangle$, is the continuum part of the damaging state (channel), $|0\rangle$. The sink, $|S_4\rangle$, is the continuum part of the CTS (channel), $|4\rangle$. All matrix elements, $V_{mn}$, in the Hamiltonian (\ref{Ham}) describe the interactions between the discrete excitonic states. The parameters, $\Gamma_0$  and $\Gamma_4$ (see below),  characterize the rates of escape from the discrete states,  $|0\rangle$ and $|4\rangle$, to their corresponding sinks.  Note, that the dynamics of escape from a discrete energy level to its associated continuum can be described by a single rate, $\Gamma_0$ (or $\Gamma_4$),  only in the approximation of an infinitely wide band of the sink. (The Weisskopf-Wigner model \cite{SM,WW}.)  For narrow or intermediate band of the sink, the dynamics of escape to the continuum is more complicated, and cannot be described by a single rate. (See, for example, \cite{Bennett,Ber6}.) But for the  purposes of this paper, we will use the Weisskopf-Wigner approximation for both sinks, assuming that for narrow or intermediate band of the sink, a single average (or renormalized) escape rate could be used. 

In our model, the dimer shown in Fig. \ref{S2} can, in principle, interact with the site, $|3\rangle$, in two different ways: (1) Both matrix elements, $V_{13}$ and $V_{23}$ (not shown in Fig. \ref{S2}) are active (non zero). In this case, both chlorophylls, $A_5$ and $B_5$, of the dimer interact  with the site, $|3\rangle$. (2) The matrix element, $V_{23}=0$, so only  $ChlA_5^*$ interacts with the site, $|3\rangle$.  Note, that for the CP29 minor complex, case (2) is more realistic as, according to \cite{Ahn},  $ChlB_5$ is located further away from $Zea$ ($\sim 13\AA$) than is $ChlA_5$ ($\sim 6\AA$). For simplicity, we use case (2) in our numerical simulations. But consideration of case (1) does not present any difficulties. 

An important part of our model is the stochastic protein environment. We use an interaction of the system with the environment in the form of   ``diagonal" stationary telegraph noise, $\xi(t)$, which acts  on all five {\it discrete} states, $|n\rangle$, and does not include ``non-diagonal" components. Note, that our model allows a generalization by considering  independent noises acting on different excitonic sites \cite{Nest4}. We also assume that the interaction constants with noise can be different at all five sites (discrete energy levels).  However, if one chooses equal  constants of interaction with noise, $\xi(t)$, at all five different sites, the effect of noise disappears. (The same occurs if one uses  a thermal bath instead of noise.) This does not happen if independent noises act on different sites \cite{Nest4}. 

Here we would like to make three comments. First,  diagonal noise (or a diagonal thermal bath) was previously  used (with some conditions) in the Marcus theory \cite{Marcus1,Xu}, and in many of its generalizations, including the description of coherent quantum effects in ET in LHCs, mentioned above. Note, that the interaction with diagonal noise leads to the relaxation processes of the ET indirectly--only through the matrix elements, $V_{nm}$ in the Hamiltonian, below. On the other hand, a non-diagonal noise (or a non-diagonal interaction with the thermal bath) leads also to {\it direct} relaxation processes \cite{Ber5}. Second, the effects of non-diagonal noise (or a non-diagonal interaction with the thermal bath) could be important under some conditions \cite{Ber5,For,Nal}, which we do not discuss here (as they are believed to reveal themselves on time-scale larger than the primary ET processes considered here). Third, usually the protein environment is modeled by a thermal bath, instead of  external (classical) noise \cite{Marcus1,Xu}. The main difference is that the long-time asymptotic leads for a thermal bath to a Gibbs distribution (in the diagonal representation of the system Hamiltonian). At the same time, for noise,  the long-time asymptotic leads to equal distribution. (See, for example, \cite{Nes1,Nes2,Ber6}, and references therein.) Because room temperature corresponds to,  $T\approx 25meV$, the difference between these two asymptotics becomes significant for the redox potential (distance between the donor and acceptor energy levels), $\epsilon> 25meV$.  As mentioned above, there are also the hybrid approaches which allow one to model a thermal environment using stochastic processes \cite{Moix,Abramavicius}. We use an approach based on a noisy environment mainly for the following three reasons: 
 First, we believe that for our purposes the noisy environment is an adequate approach, since the NPQ mechanism in our model leads to rapid saturation (stationary regime), which usually reveals itself on a time-scale less than the time of equilibration in the system with only discrete electron energy levels. Second, as mentioned above, it should be expected that most bio-systems operate in the functional mode (including the  NPQ),  in noisy not thermal environment, or at least, in combination. Third, the use of random telegraph noise allows us to derive an {\it exact} system of linear ordinary differential equations for the reduced density matrix, which describes the NPQ process in our model. (For an approach based on a thermal bath at room temperature, with a relatively strong constant of interaction between the system and the bath, the exact system of equations is not known \cite{Xu}. Moreover, a controlled perturbation theory also does not exist at present.) This is important for our analysis, as our main intention is to describe the dynamics of the NPQ process, which is much easier to do by using an exact system of equations.

\section{Mathematical formulation}
\subsection{The Hamiltonian}
Generally, we consider the system with $N$ discrete energy levels (excitonic sites), with each level coupled to its independent sink. The corresponding time-dependent Hamiltonian has the form,
\begin{eqnarray}
\label{Ham}
\mathcal H(t) = \sum^N_{n=1} E_n|n\rangle\langle n| + \sum_{m,n} \beta_{mn}(t)|m\rangle\langle n|  \nonumber \\
 +  \sum^{N}_{n=1}\sum^{N_n}_{i_n=1} ( E_{i_n}|i_n\rangle\langle  i_n | +V_{ni_n}|n\rangle\langle  i_n | + V_{i_n n}|i_n\rangle\langle n |\big),
\end{eqnarray}
where, $m,n = 1,2, \dots,N$. Each site, $n$, has the exciton energy, $E_n$. The functions, $\beta_{mn}(t)$, generally describe the time-dependent matrix elements between different sites, noise, and possible regular time-dependent forces. The matrix elements, $E_{i_n}$ and $V_{i_nn}$, describe the energy levels of the sink attached to the site, $n$, and the interaction between  the site, $n$, and the attached sink, correspondingly. We assume that the energy levels of all sinks are sufficiently dense, so one can perform an integration instead of a summation over the index, $i_n$. Then, we have,
\begin{eqnarray}\label{ah1}
\mathcal H(t) = &\sum_{n}E_n|n\rangle\langle n| + \sum_{m,n}\beta_{mn}(t)|m\rangle\langle n +\sum_{n}\int E|E\rangle\langle E|g_n(E)dE \nonumber \\
&+\sum_{n} \Big( \int\alpha_n(E)|n\rangle\langle E|g_n(E) dE + \rm h.c.\Big) 
 \end{eqnarray}
where, $g_n(E)=dn/dE$, is the density of electron states, and $V_{ni_n}\rightarrow \alpha_n(E)$.

 \subsection{The non-Hermitian Hamiltonian and the Liouville equation}
 
 Using the standard Feshbach projection method \cite{SM,RI,RI3,VZ}, one can show that in the Weisskopf-Wigner pole approximation \cite{SM,WW,scully} the $N$-level system interacting with $N$ independent sinks can be described by the effective non-Hermitian Hamiltonian,  $ \tilde{\mathcal H}= {\mathcal H}- i \mathcal W$, where,
 \begin{eqnarray}\label{DM1}
 {\mathcal H} = \sum_{n}\varepsilon_n |n\rangle \langle n|  +\sum_{m, n} \beta_{mn}(t) |m\rangle \langle n|,
 \end{eqnarray}
is the dressed Hamiltonian, and the non-Hermitian part is,
 \begin{eqnarray}
 \label{Gamma}
{\mathcal W} = \frac{1}{2}\sum_{n}\Gamma_n|n\rangle \langle n|.
 \end{eqnarray}
 In (3), $\epsilon_n$ is the renormalized energy, $E_n$, of the state, $|n\rangle$,	
 	and the parameter, $\Gamma_n$, in (\ref{Gamma})  is the tunneling rate to the $n$-{th} sink, 
 ${\Gamma_n} = 2\pi|\alpha_n(E_n)|^2g_n(E_n)$ (see Appendix A).

The dynamics of the system can be described by the Liouville  equation (see Appendices A and B),
 \begin{eqnarray}\label{DM1}
    \dot{ \rho} = i[\rho,\mathcal H] - \{\mathcal W,\rho\},
 \end{eqnarray}
 where $\{\mathcal W,\rho\}= \mathcal W\rho +\rho\mathcal  W$.

We define the ET efficiency of tunneling to all $N$ sinks as,
\begin{eqnarray}\label{ET1} 
\eta(t) = 1 - {\rm Tr}(\rho(t)) =  \int_0^t {\rm Tr}\{\mathcal W,\rho(\tau)\} d \tau.
\end{eqnarray}
This can be expressed as the sum of time-integrated probabilities of trapping an electron into the $n$-th sink \cite{RMKL,CDCH}, 
\begin{eqnarray}
\eta_n(t) = \Gamma_n \int_0^t \rho_{nn}(\tau)d \tau,~ (\eta(t)=\sum_{n=1}^N\eta_n(t)).
\label{Eq16ar}
\end{eqnarray}
We call, $\eta_n(t)$ ($0\le \eta_n(t)\le 1$) the efficiency of the $n$-th sink. In particular, for our five-site model, presented in Fig. (\ref{S2}), a complete suppression of the damaging channel {\it at all times} means that both, $\eta_0(t)=0$ and $\rho_{00}(t)=0$ (the probability of finding the electron in the  state, $|0\rangle$).

\section{Noise-assisted ET in the NPQ mode}
\label{sec:2}

In the presence of  noise, the quantum dynamics of the ET can be described by the following effective non-Hermitian Hamiltonian (for details see Appendix B),
\begin{eqnarray} \label{Eq2b}
 \tilde{\mathcal H}= \sum_n \varepsilon_n |n\rangle\langle  n |+ \sum_{m,n} \lambda_{mn}(t)|m\rangle\langle  n |
 + \sum_{m \neq n} V_{mn} |m\rangle\langle  n | \nonumber \\
 - i\frac{\Gamma_0}{2} |0\rangle \langle 0| - i\frac{\Gamma_4}{2} |4\rangle \langle 4|, \quad m,n = 0,1,\dots 4,
\end{eqnarray}
where, $\lambda_{mn}(t)$, describes the influence of noise.  The diagonal matrix elements of noise, $\lambda_{nn}(t)$, are responsible for decoherence. Then, the relaxation in this case occurs only if $V_{mn}\ne 0$. The off-diagonal matrix elements, $\lambda_{mn}(t)$ ($ m \neq n$), lead to {\it direct} relaxation processes.

Below we restrict ourselves to consider only the diagonal noise effects. Then, one can write, $\lambda_{mn}(t) = \lambda_n \delta_{mn}\xi(t) $,  where $\lambda_n$ is the interaction constant at site, $n$.  We consider stationary telegraph noise to be described by the random variable,  $\xi(t) = \zeta(t) -\bar{\zeta} $, so that,
 \begin{eqnarray}\label{chi_8}
&\langle \xi(t)\rangle =0, \\
 &\chi(t-t')=\langle \xi(t)\xi(t')\rangle,
\end{eqnarray}
where, $\chi(t-t') = \sigma^2 e^{-2\gamma |t-t'|}$, is the correlation function of noise. The average value, $\langle \zeta(t)\rangle =\bar{\zeta}$, is included in the renormalization of the excited electron energy at each site, $n$, in Eq. (\ref{Eq2b}) as: $\varepsilon_n\rightarrow \varepsilon_n+d_n\bar{\zeta}$.

Writing the effective non-Hermitian Hamiltonian as, $ \tilde{\mathcal H}= {\mathcal H}- i \mathcal W$, where,
\begin{eqnarray}
\mathcal W =\frac{\Gamma_0}{2} |0\rangle \langle 0| +\frac{\Gamma_4}{2} |4\rangle \langle 4|,
\end{eqnarray}
we find that the evolution of the  average  components of the density matrix is described by the following closed system of ordinary differential equations (see Appendix B):
\begin{eqnarray} \label{IB4}
\frac{d}{dt}{\langle{\rho}}\rangle =&i[\langle\rho\rangle,\mathcal H] - \{\mathcal W,\langle\rho\rangle\} - i B\langle\rho^\xi\rangle, \\
\frac{d}{dt}{\langle{\rho^\xi}}\rangle =&i[\langle\rho^\xi\rangle,\mathcal H] - \{\mathcal W,\langle\rho^\xi\rangle - i B\langle\rho\rangle- 2\gamma \langle\rho^\xi\rangle ,
\label{IB5}
\end{eqnarray}
where $B = \sum_{m,n}(d_m - d_n)|m\rangle \langle n|$ and $d_{m}=\lambda_m \sigma$. We set $\langle\rho^\xi\rangle = \langle\xi\rho \rangle/\sigma$. The average,  $\langle \,\dots \,\rangle$, is taken over the random process.

By manipulating the rates, $\Gamma_{0,4}$, of tunneling into the sinks, one can modify the dynamics of the whole system. 
 The approach based on sinks, has been used to model the ET dynamics in photosynthetic complexes. (See, for example,  \cite{Pudlak1,Pudlak2,PPM,PPM1}, and references therein.) Below, we will use this approach to model a possible role of quantum interference effects and sinks in the NPQ mechanism.

\section{Characteristics of an LHC dimer}
The importance of the different LHC dimers based on chlorophyll molecules in the ET in LHCs and RCs is widely recognized \cite{Harrop,Dreuw1,Cheng,Ahn,Val,Duffy1,Duffy2,Renger,Nov2,Muh}. (See also references therein.)
Following \cite{Dreuw1,Cheng,Ahn}, we will assume that in the NPQ mode, in the photosynthetic complex (we choose, CP29, for concreteness)  an LHC dimer is realized consisting of the excited states of two neighboring chlorophylls, $ChlA_5^*$ and $ChlB_5^*$. To characterize the states of the {\it isolated} LHC dimer, we consider in this sub-section in the Hamiltonian (\ref{Ham}) only two states, $|1\rangle$ and $|2\rangle$, and the interaction between them,
\begin{equation}
\label{Hd}
H_{dimer} =E_1|1\rangle\langle 1|+E_2|2\rangle\langle 2|+(V_{12}|1\rangle\langle 2|+h.c.).
\end{equation}
This dimer has two orthogonal eigenstates, which in the basis of the site states, 
\begin{equation}
\label{one}
|1\rangle=\left ( \begin{array}{c}
1 \\
0
\end{array}
\right),~
|2\rangle=\left ( \begin{array}{c}
0 \\
1
\end{array}
\right ),
\end{equation}
have the form,
\begin{equation}
\label{phim}
 |\varphi_{-} \rangle= \frac{2|V_{12}|}{\sqrt{(\delta + \sqrt{\delta^2 +4|V_{12}|^2 })^2 +4|V_{12}|^2}}
\left ( \begin{array}{c}
 1 \\
 -\frac{\delta + \sqrt{\delta^2 +4|V_{12}|^2 }}{2V_{12}}
 \end{array}
 \right ) ,
 \end{equation}
 with the energy eigenvalue, $E_-=(E_1+E_2)/2-\sqrt{\delta^2+4|V_{12}|^2}/2$, and
 \begin{equation}
 \label{phip}
 |\varphi_{+} \rangle= \frac{2|V_{12}|}{\sqrt{(\delta + \sqrt{\delta^2 +4|V_{12}|^2 })^2 +4|V_{12}|^2}}
\left (\begin{array}{c}
 \frac{\delta + \sqrt{\delta^2 +4|V_{12}|^2 }}{2V_{21}}\\
 1
 \end{array}
 \right ),
 \end{equation}
 with the energy eigenvalue, $E_+=(E_1+E_2)/2+\sqrt{\delta^2+4|V_{12}|^2}/2$, where, $\delta=E_1-E_2$.  
 
 Here we would like to note, that the eigenstates of the dimer depend significantly on the sign of the matrix element, $V_{12}$, of the interaction between two chlorophylls \cite{Duffy1,Duffy2}. To see this, suppose, for simplicity, that $\delta=0$, so that the unperturbed excited energy levels of both chlorophylls are degenerate, $E_1=E_2=E$. As  follows from (\ref{phim}) and (\ref{phip}), in this case, $|\varphi_-\rangle=(1/\sqrt{2})(|1\rangle-sign({V_{12})|2\rangle})$, and $|\varphi_+\rangle=(1/\sqrt{2})(sign({V_{12})|1\rangle+|2\rangle})$. Then, in principle, both $V_{12}>0$ and $V_{12}<0$ 
 situations can be realized in natural LHCs \cite{Duffy2}. Indeed, suppose that the excitons associated with the excited states of two chlorophylls interact mainly through the dipole-dipole interaction. Then, the geometries of the chlorophyll molecules and the orientations of their transitional dipole moments become important. In particular, in the degenerate case, $\delta=0$, when $V_{12}>0$, the antisymmetric eigenstate, $|\varphi_-\rangle=(1/\sqrt{2})(|1\rangle-|2\rangle)$, is the ground state of the dimer, with the eigenenergy, $E_-=E-|V_{12}|$.  When $V_{12}<0$, the symmetric eigenstate, $|\varphi_-\rangle=(1/\sqrt{2})(|1\rangle + |2\rangle)$, is the ground state of the dimer, with the same energy. Note, that the interaction of the two chlorophylls, which form the dimer, can have a rather complicated dependence on the distance between them. Namely, the F\"orster dipole-dipole type of interaction for relatively large distances ($\gtrsim 1nm$) can be modified for the Dexter mechanism for small distances ($\lesssim 1nm$), which involves the electron exchange interaction \cite{OEVERING}. 
 
 When the two chlorophylls, $A_1$ and $A_2$, form the strongly coupled dimer, so that $|V_{12}/ \delta|\gtrsim 1$, the parameter, $\delta$, can be small, but the matrix element of their interaction can become large enough, if the chlorophylls are near each other. In this case, $\Delta E\approx 2|V_{12}|$. This sort of the chlorophyll dimer is realized in the so-called H-aggregate quasi-parallel molecules. The upper-lying state, $|\varphi_+\rangle$, is called the ``bright" (``dipole-allowed") state, and the low-lying state, $|\varphi_-\rangle$, is called the ``dark" (``dipole-forbidden") state that  traps the exciton \cite{Duffy2}. 
 
 \subsection{Weakly and strongly coupled  LHC dimers}  
 Introduce the parameter,
 \begin{equation}
 \label{dimer}
 \mu_d=|V_{12}/\delta|.
 \end{equation}
 It is easy to see that when $\mu_d\ll 1$, both eigenstates, (\ref{phim}) and  (\ref{phip}), become close to the unperturbed states (\ref{one}) of two separated chlorophylls. In this case, the LHC dimer is called ``weakly coupled". The LHC dimer is called ``strongly coupled" when,  the value of $\mu_d$ is not too small. We can say that the LHC dimer is strongly coupled if, $\mu_d\gtrsim 1$.
 
 As was indicated in \cite{Ahn},  the dihedral angle between $ChlA_5$ and $ChB_5$ is $\approx 41^0$.   So, we assume in the following that $V_{12}>0$. In this case, the energy level of the eigenstate, $|\varphi_{+}\rangle$, is located above the energy level of the eigenstate, $|\varphi_{-}\rangle$, with an energy gap between them, $\Delta E=E_+-E_-=\sqrt{\delta^2+4|V_{12}|^2}$. 
 
 For the CP29 antenna complex, the exciton Hamiltonian is presented in \cite{Muh}, for the non-NPQ mode. In particular, the LHC dimer based on two chlorophylls, $(Chla604-Chlb606)^*$, was specified: $\varepsilon_{a604}\approx 1.839eV$, $\varepsilon_{b606}\approx 1.938eV$; $V_{a604,b606}\approx 8.3meV$. So, in this case, the energy gap between these two chlorophylls is: $\delta=\varepsilon_{a604}-\varepsilon_{b606}\approx -99meV$. For this dimer, the value of the parameter $\mu_d$ is: $\mu_d=|\delta/V_{a604,b606}|\approx 0.08\ll1$. So, this dimer should be considered to be weakly coupled. (Note that the heterodimers based on two chlorophylls, $A$ and $B$, are also weakly coupled in many other bio-systems, not only in the CP29 antenna complex.) 
 
 However, below, in our numerical simulations, we will consider both weakly and strongly coupled cases for the dimer, $(ChlA-ChlB)^*$, in Fig. \ref{S2} (shown in the rectangle). The main reason is that in the NPQ mode, when the protein environment experiences conformational modifications, the characteristics of the dimer, $(ChlA-ChlB)^*$, can also change. Indeed, the reconstruction energy and the interaction matrix element, $V_{12}$, could increase significantly. Also, because of protonation in the NPQ mode, the local electric fields on the sites can change, resulting in modification of the redox potential. Then, the weakly coupled dimer could become a strongly coupled dimer. Additional experimental efforts are required in order to determine when this occurs. 
 
 \subsection{Remarks on the initial conditions}
  Both above states,  $|\varphi_{+}\rangle$ and $|\varphi_{-}\rangle$, are  eigenstates only of the isolated dimer. When the dimer is formed, its interaction with sunlight should include an additional term in the Hamiltonian (\ref{Hd}): $H_{p} =-\hat d\vec{E}$, where $\hat d$ is the operator of the optical dipole transition, and $\vec{E}$ is the electric field of sunlight. Then, in frequency units, this additional term is proportional the the Rabi frequency, $\Omega_R=dE$. In this case, the above states, $|\varphi_{+}\rangle$ and $|\varphi_{-}\rangle$, are not eigenstates of the system.  Also, this dimer interacts in real systems with other sites, which makes the Hamiltonian (\ref{Hd}) even more complicated. So, the dark state, $|\varphi_{-}\rangle$, can be significantly populated. Another way to populate the dark state is through the non-radiative transition: $|\varphi_{+}\rangle\rightarrow |\varphi_{-}\rangle$ \cite{Duffy1}.  

 In the numerical simulations below, when populating initially the LHC dimer, $(ChlA-ChlB)^*$, we mainly use the initial condition for the wave function of the whole system shown in Fig. \ref{S2}: $|\varphi(0)\rangle=|\varphi_{-}\rangle$. In this case, when the dimer is strongly coupled, $\mu_d\gtrsim 1$, the initial wave function is delocalized over the two chlorophylls. But when the dimer is weakly coupled, $\mu_d\ll 1$, the initial wave function is mainly localized on the $ChlA^*$.
 In a few cases in the NPQ mode, we also considered that  the upper state of the LHC dimer, $|\varphi_{+}\rangle$, was initially populated. 

\section{Non-NPQ regime: Interference effects, two-level limit, influence of sink and noise}

\begin{figure}[tbh]
	\begin{center} \scalebox{0.455}{\includegraphics{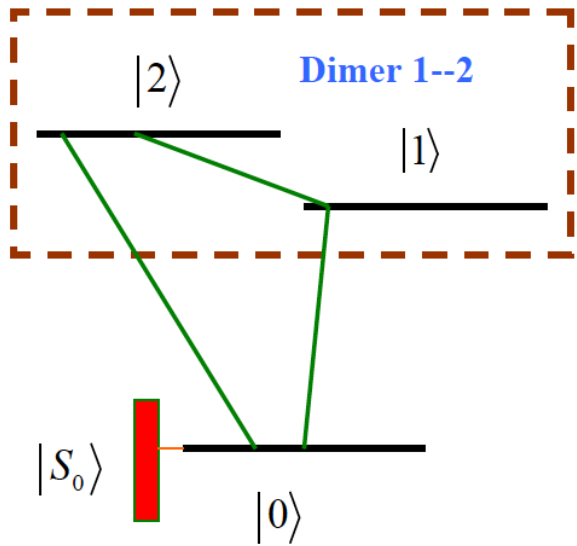}}
	\end{center}
	\caption{Schematic of a reduced model-three-level system. The rectangle indicates the initial population of the dimer. The state, $|0\rangle$, is a damaging  state; $|S_0\rangle$ is the damaging sink.
		\label{S2a}}
\end{figure}

Here we consider a simplified version of the system presented in Fig. \ref{S2}.  This reduced three-level system is shown in Fig. \ref{S2a}. It describes the interaction of the LHC dimer, $(ChlA-ChlB)^*$ (shown in the rectangle, in Fig. \ref{S2a}), with the discrete damaging level, $|0\rangle$, and its sink, $|S_0\rangle$. In numerical simulations, we use different parameters for the dimer, and different initial conditions.

We also would like to mention  here that the system shown in Fig. \ref{S2a} consists not only of the ``main" chlorophyll LHC dimer $(1-2)$, but also on the dimers, $(1-0)$ and $(2-0)$. In particular, below in sub-section 6.2, we will use the notation for the LHC dimer, $(1-0)$, similar to the ``main" LHC dimer. Namely, the parameter, $\mu_{10}\equiv|V_{10}/\delta_{10}|$ ($\delta_{10}=\varepsilon_1-\varepsilon_0$), will be used,   which characterizes  if this dimer is strong ($\mu_{10}\gtrsim 1 $) or weak ($\mu_{10}\ll 1$).

\subsection{Qualitative analysis of the interference effects}

 To understand qualitatively a possible influence of the quantum interference effects in both non-NPQ and NPQ modes (suppression of the channel, $|0\rangle$, and its sink, $|S_0\rangle$), we first consider instead of (\ref{Ham}) the following simplified Hamiltonian (see Fig. \ref{S2a}, without a sink, $|S_0\rangle$, and noise), 
\begin{equation}
\label{HS}
\bar H_3 =E_0|0\rangle\langle 0|+E_1|1\rangle\langle 1|+E_2|2\rangle\langle 2|+
(V_{12}|1\rangle\langle 2|+V_{10}|1\rangle\langle 0|+V_{20}|2\rangle\langle 0|+h.c.),
\end{equation}
with the additional condition: $E_1=E_2=E$ ($\delta=0$), and all real matrix elements, $V_{nm}\neq 0$ $(n,m=0,1,2)$. This case corresponds to a  strongly coupled dimer, $\mu_d=\infty$. We represent the wave function as,
\begin{equation}
\label{wf}
|\varphi(t)\rangle=c_0(t)|0\rangle+c_1(t)|1\rangle+c_2(t)|2\rangle.
\end{equation}
The system of equations for the complex amplitudes, $c_{0,1,2}(t)$, is,
\begin{eqnarray}
\label{sys}
i\dot c_0=E_0c_0+V_{01}c_1+V_{02}c_2,\\
i\dot c_1=Ec_1+V_{12}c_2+V_{10}c_0,\\
i\dot c_2=Ec_2+V_{21}c_1+V_{20}c_0.
\end{eqnarray}
By adding the second and the third equations, we have: $i\dot c=Ec+V_{int}c+2Vc_0$, where we use the notations: $c(t)=c_1(t)+c_2(t)$, $V_{12}=V_{21}=V_{int}$, $V_{10}=V_{20}=V=V^*$. Making the substitutions: $c(t)=A(t)e^{-i(E+V_{int})t}$, $c_0(t)=A_0(t)e^{-iE_0t}$, we have for $A(t)$ and $A_0(t)$,
\begin{eqnarray}
\label{sol}
i\dot A=2Ve^{i(E+V_{int}-E_0)t}A_0,\\
i\dot A_0=Ve^{i(E+V_{int}-E_0)t}A.
\end{eqnarray}

\begin{figure}[tbh]
	\begin{center}
		\scalebox{0.285}{\includegraphics{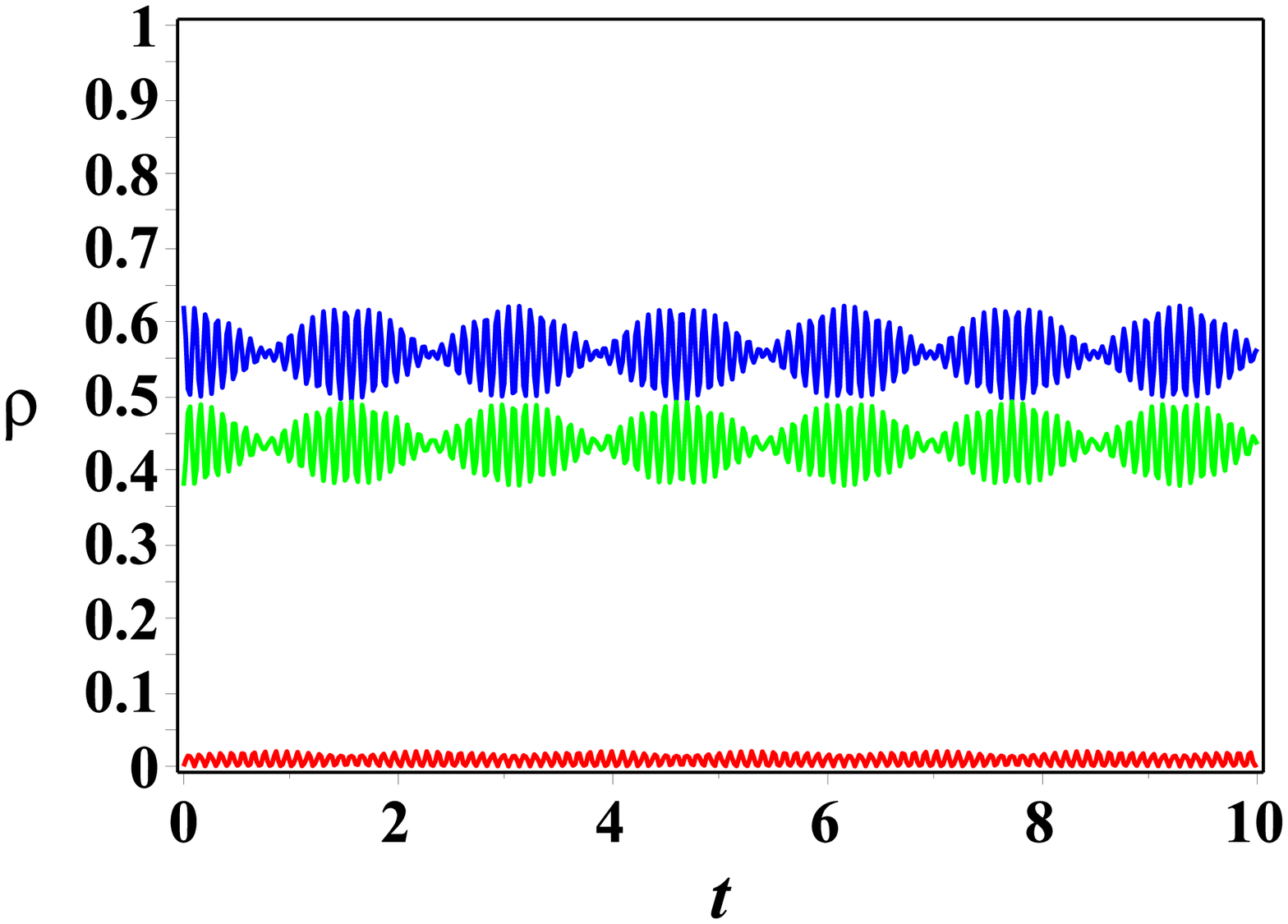}}
		(a)
		\scalebox{0.285}{\includegraphics{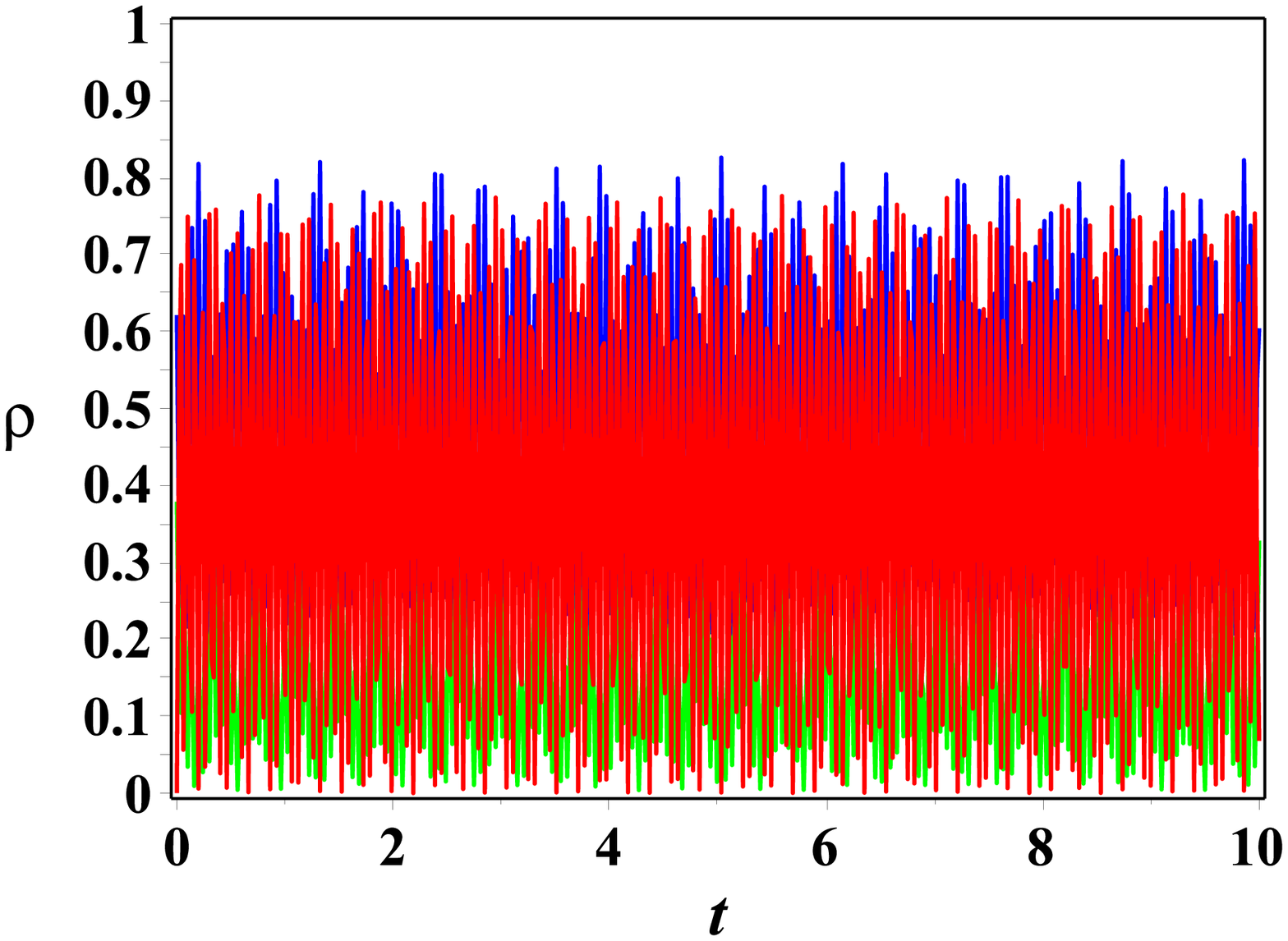}}
		{(e)}
		\scalebox{0.285}{\includegraphics{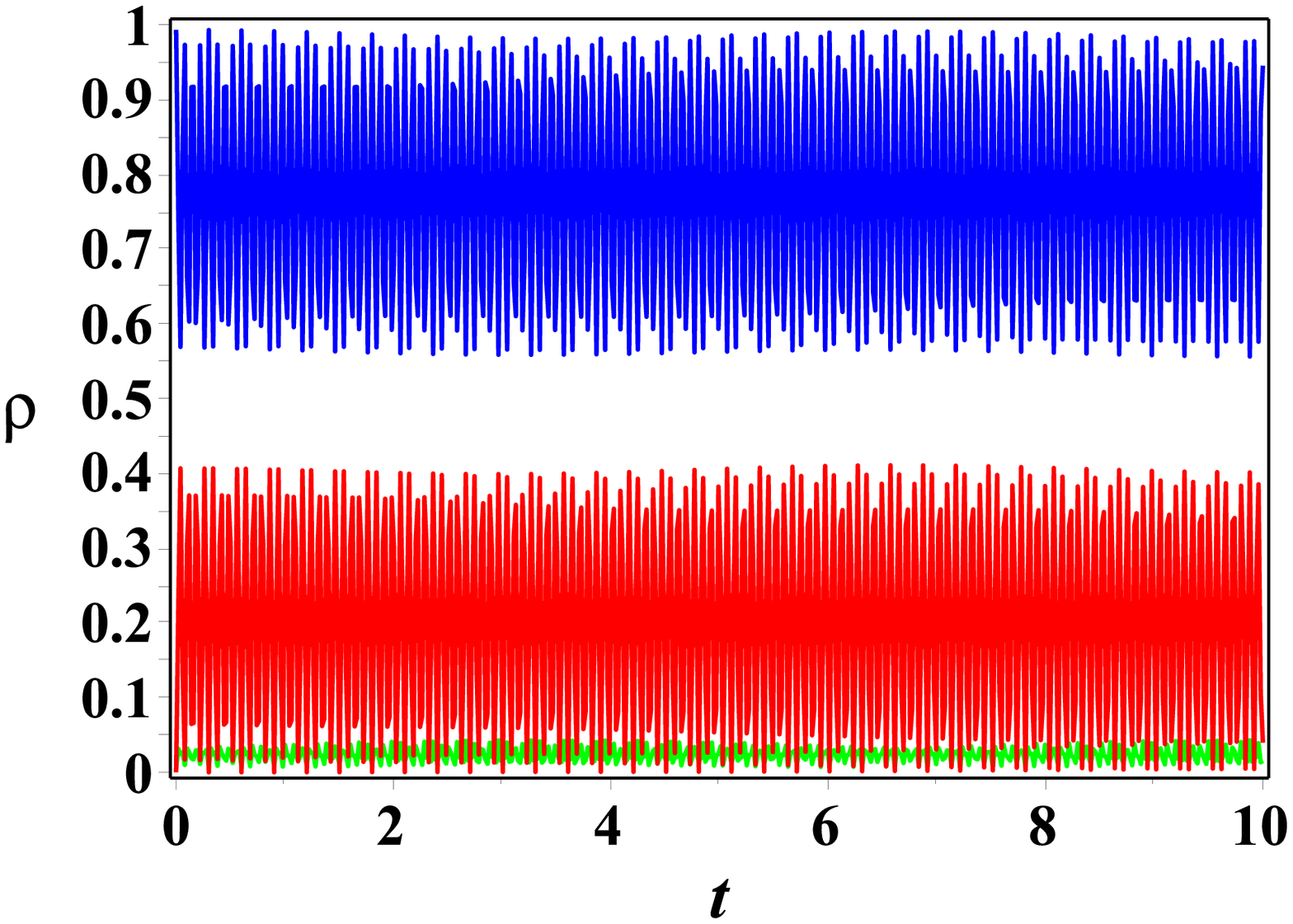}}
		(b)
		\scalebox{0.285}{\includegraphics{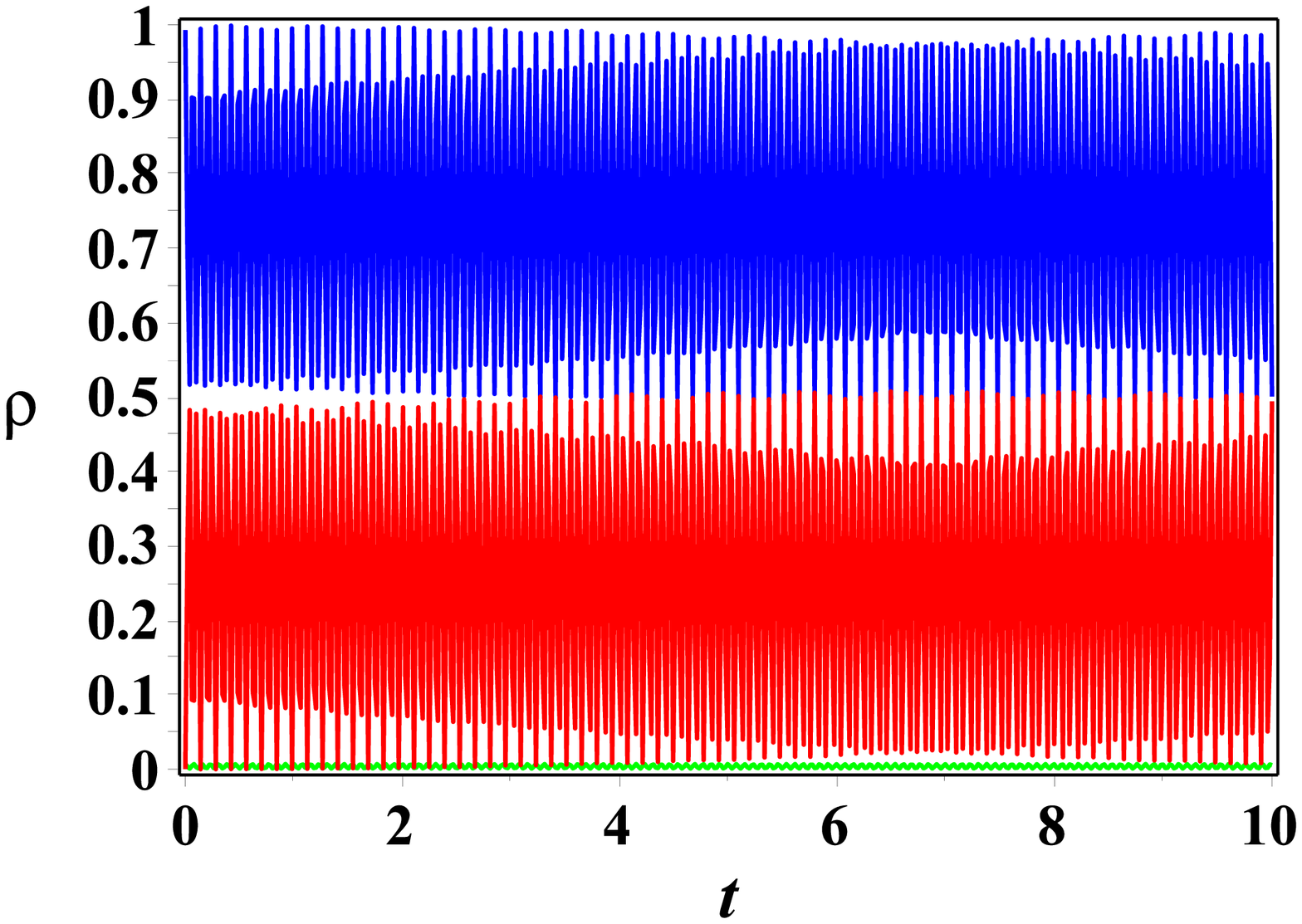}}
		{(f)}
		\scalebox{0.285}{\includegraphics{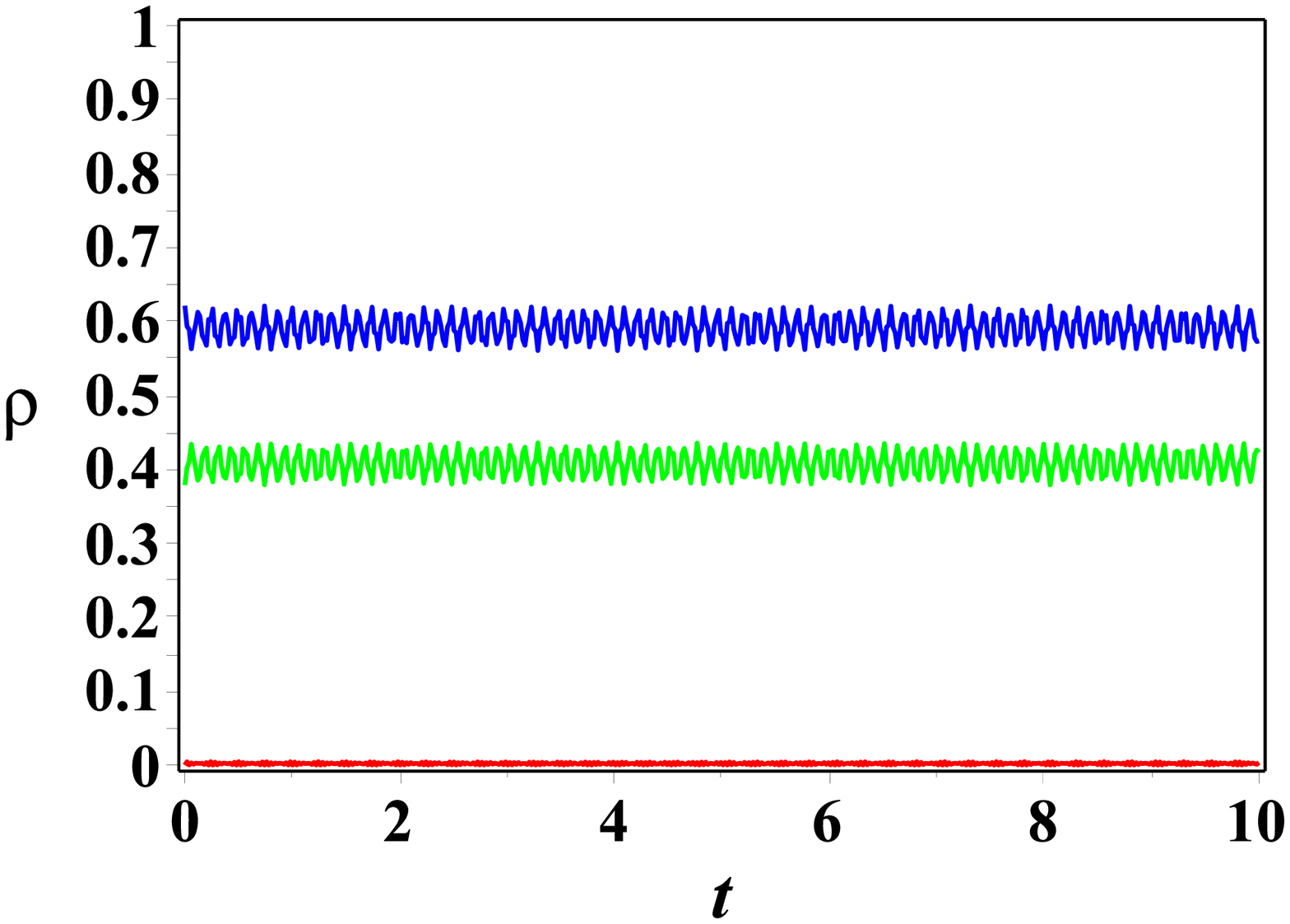}}
		(c)
		\scalebox{0.295}{\includegraphics{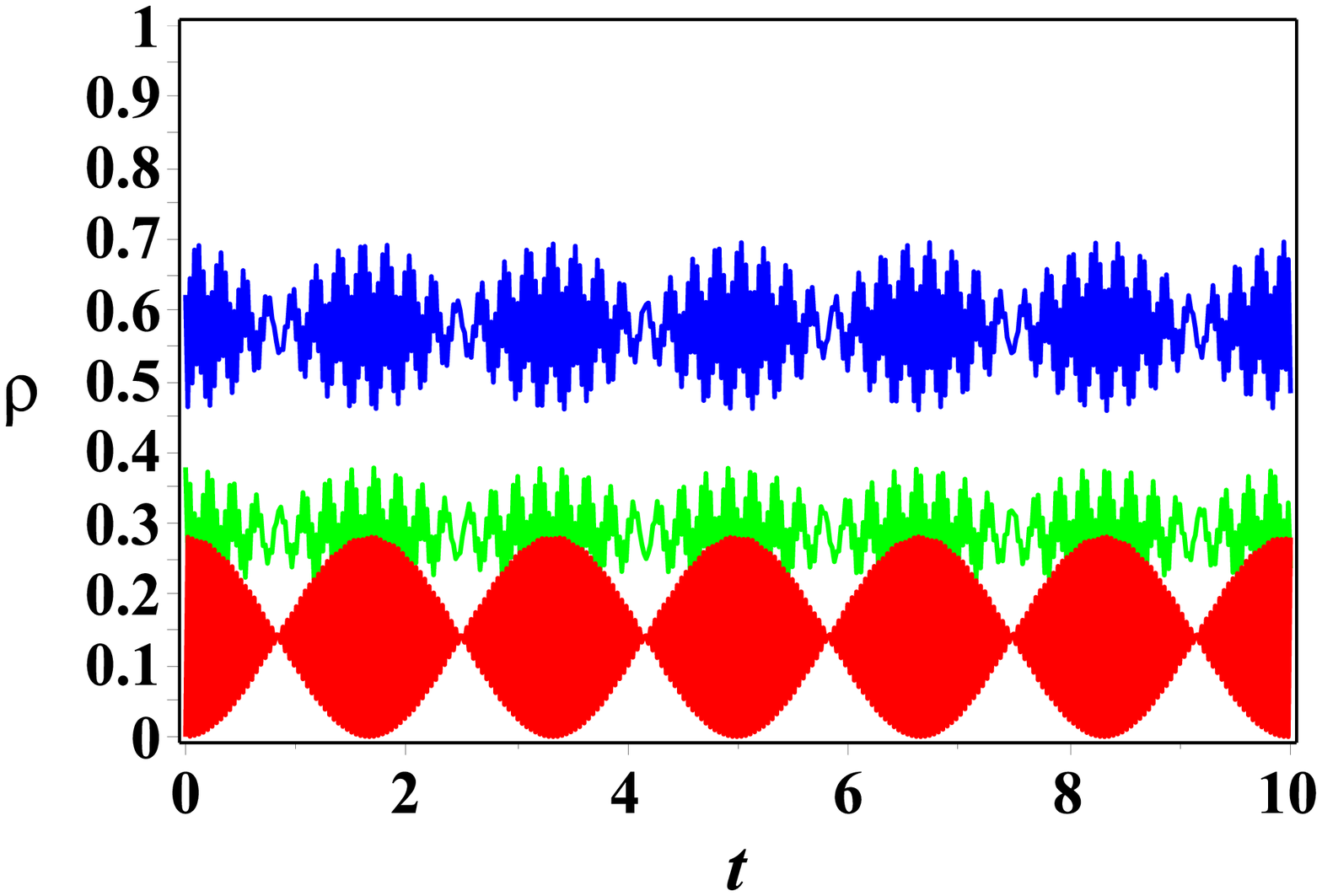}}
		{(g)}
		\scalebox{0.285}{\includegraphics{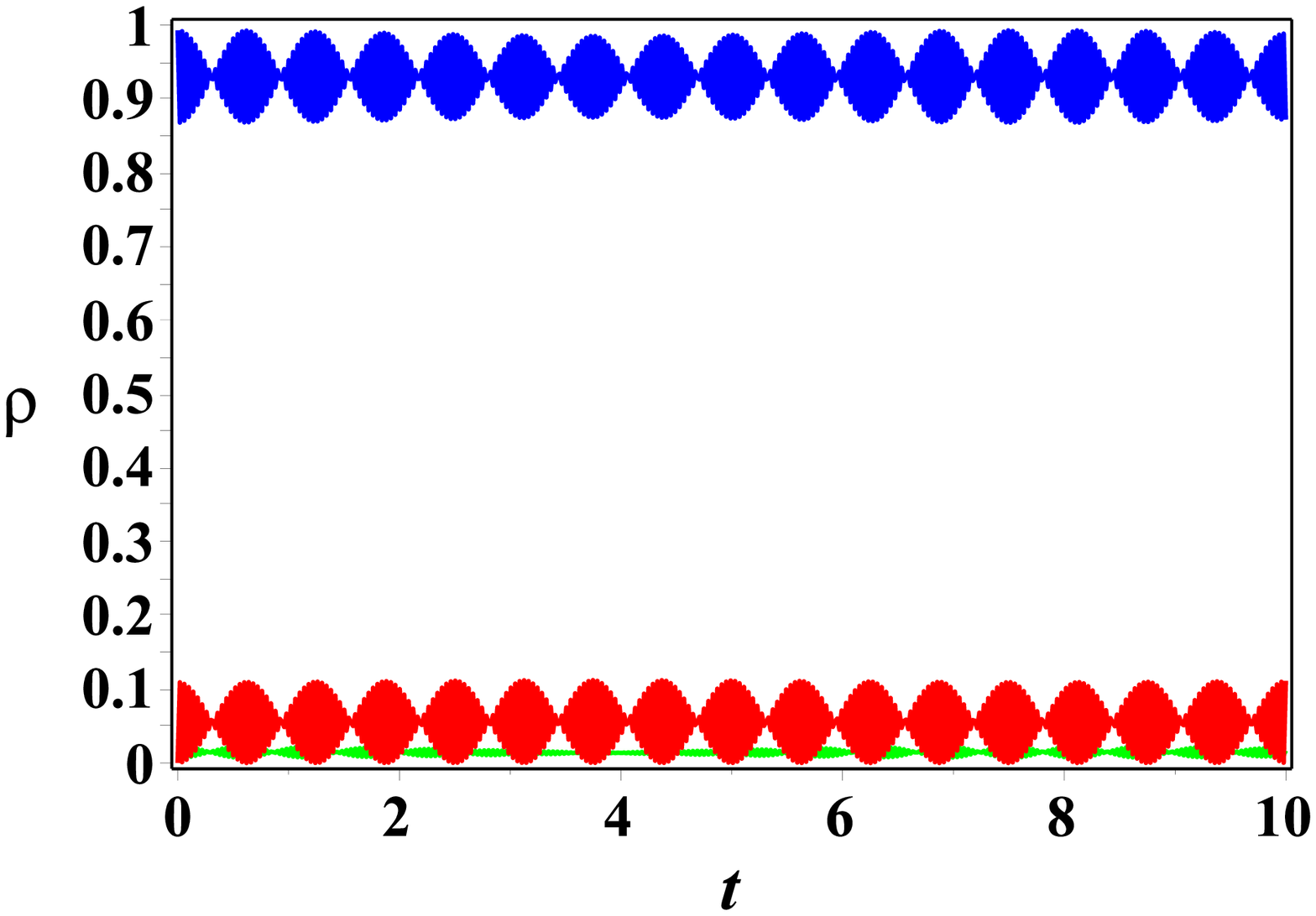}}
		(d)
		\scalebox{0.29}{\includegraphics{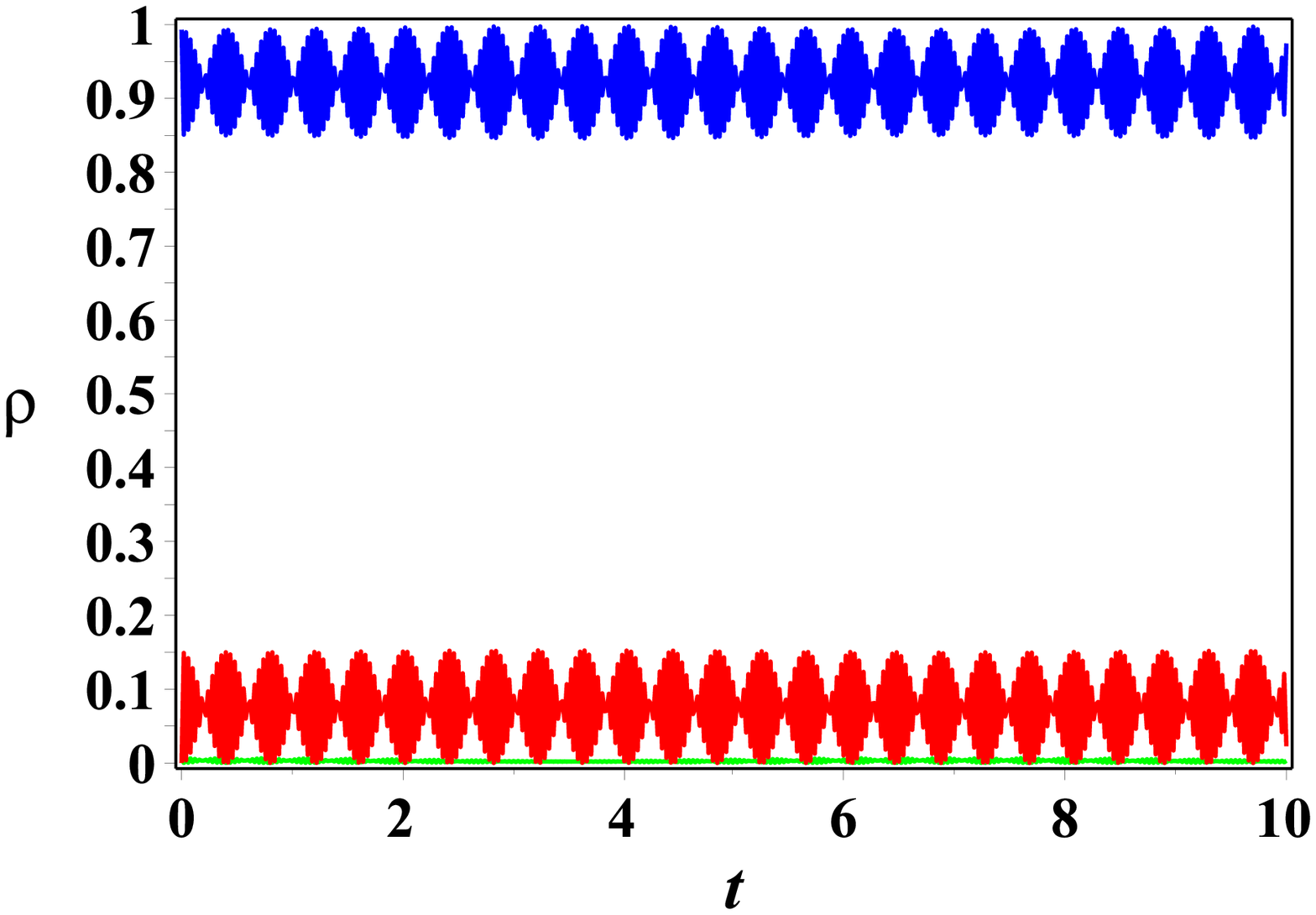}}
		{(h)}
	\end{center}
	\caption{(Color online) Three-level system, presented in Fig. \ref{S2a}, without sink and noise. Time dependence (in ps) of the density matrix components: $\rho_{00}(t)$ (red), $\rho_{11}(t)$ (blue), $\rho_{22}(t)$ (green). Parameters: $V_{10}=V_{20}=30, \varepsilon_1=60$;~(a) $V_{12}=20$, $~\varepsilon_0=0$, $\delta = -10$;~(b) $V_{12}=12,6$, $~\varepsilon_0=0$, $\delta = -150$; (c) $V_{12}=20$, $~\varepsilon_0=-90$, $\delta = -10$; (d) $V_{12}=12,6$, $~\varepsilon_0=-90$, $\delta = -150$. Initial conditions: $|\varphi(0)\rangle=|\varphi_-\rangle$. This corresponds to: $\rho_{11}(0)=0.621,~\rho_{22}(0)=0.379,~\rho_{12}=\rho_{21}=-0.485$~(a,c); $\rho_{11}(0)=0.993,~\rho_{22}(0)=0.007,~\rho_{12}=\rho_{21}=-0.083$~(b,d). Left: $V_{10}= V_{20}$. Right: $V_{10}= -V_{20}$.
		\label{A}}
\end{figure}

 \begin{figure}[tbh]
 	\begin{center}
 		\scalebox{0.285}{\includegraphics{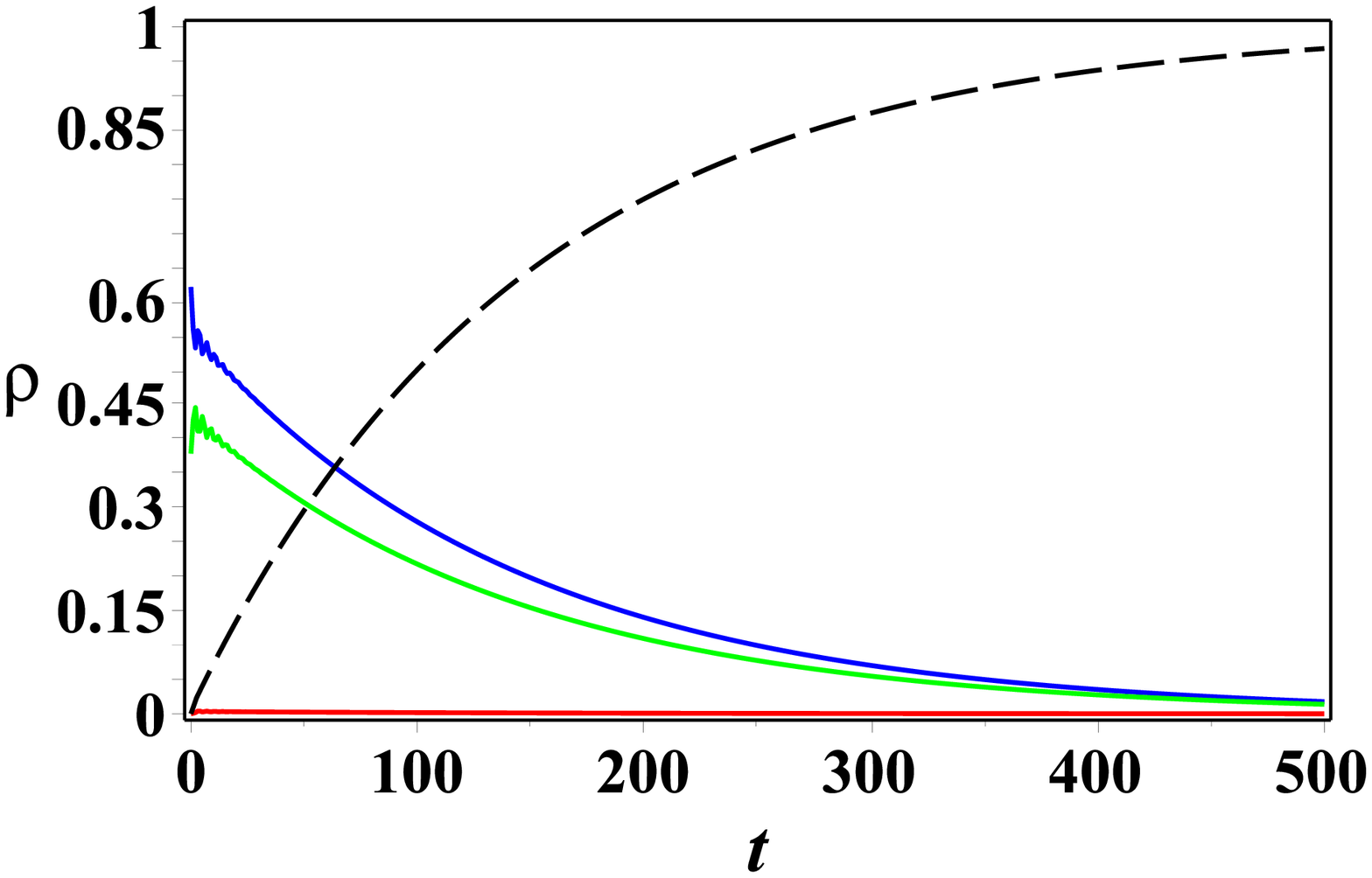}}
 		(a)
 		\scalebox{0.2925}{\includegraphics{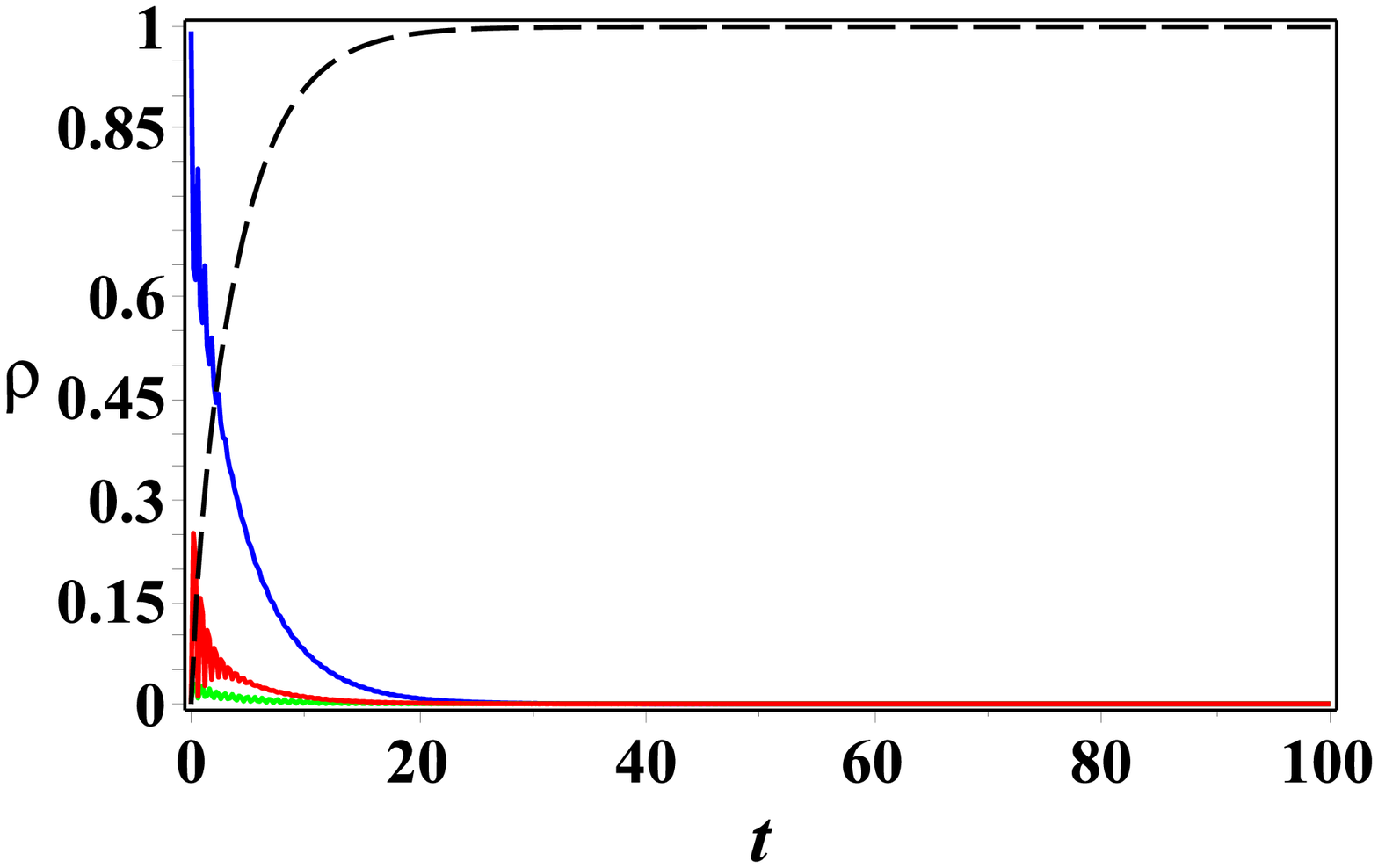}}
 		(b)
 		\scalebox{0.285}{\includegraphics{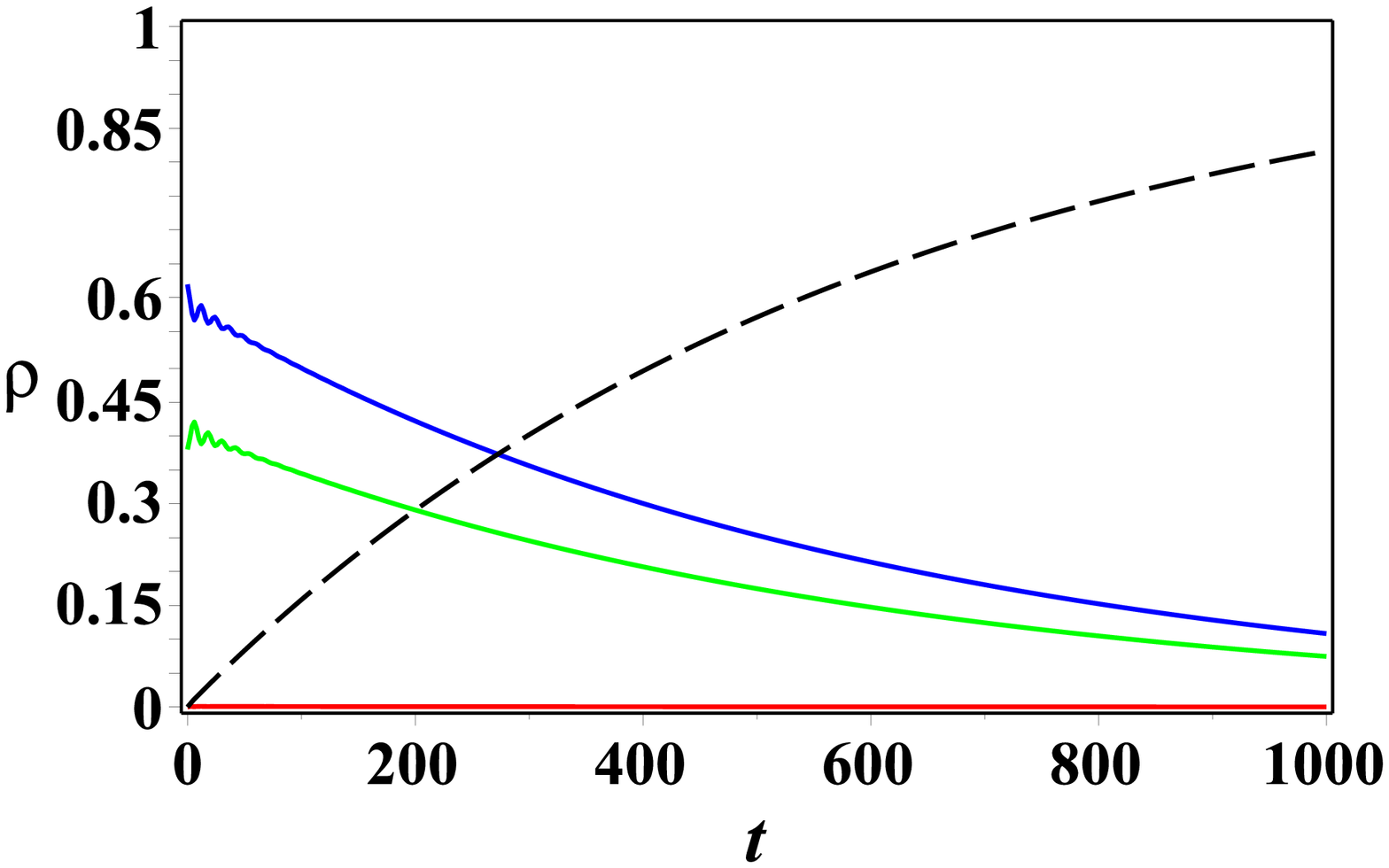}}
 		(c)
 		\scalebox{0.2925}{\includegraphics{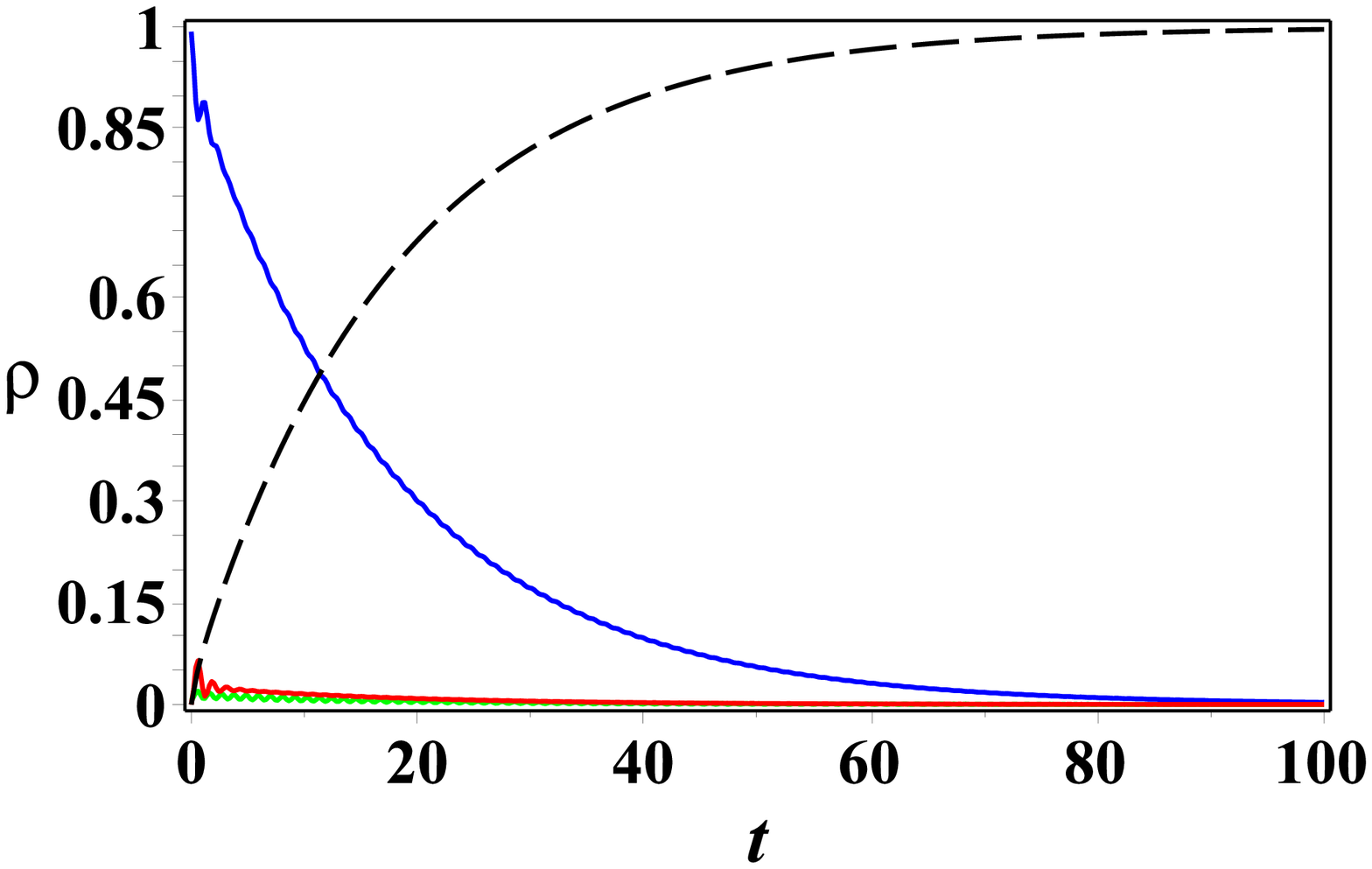}}
 		(d)
 	\end{center}
 	\caption{(Color online) Three-level system. Time dependence (in ps) of the efficiency of sink, $\eta_0(t)$ (dashed lines), and the density matrix components: $\rho_{00}(t)$ (red curves), $\rho_{11}(t)$ (blue curves), $\rho_{22}(t)$ (green curves). Parameters: $V_{10}=V_{20}=30, \varepsilon_1=60$, $\Gamma_0 =2$; (a) $V_{12}=20$, $~\varepsilon_0=0$, $\delta = -10$; (b) $V_{12}=12,6$, $~\varepsilon_0=0$, $\delta = -150$; (c) $V_{12}=20$, $~\varepsilon_0=-90$, $\delta = -10$; (d) $V_{12}=12,6$, $~\varepsilon_0=-90$, $\delta = -150$.  Initial conditions: $|\varphi(0)\rangle=|\varphi_-\rangle$. 
 		\label{A4}}
 \end{figure}
 
 \begin{figure}[tbh]
 	\begin{center}
 		\scalebox{0.275}{\includegraphics{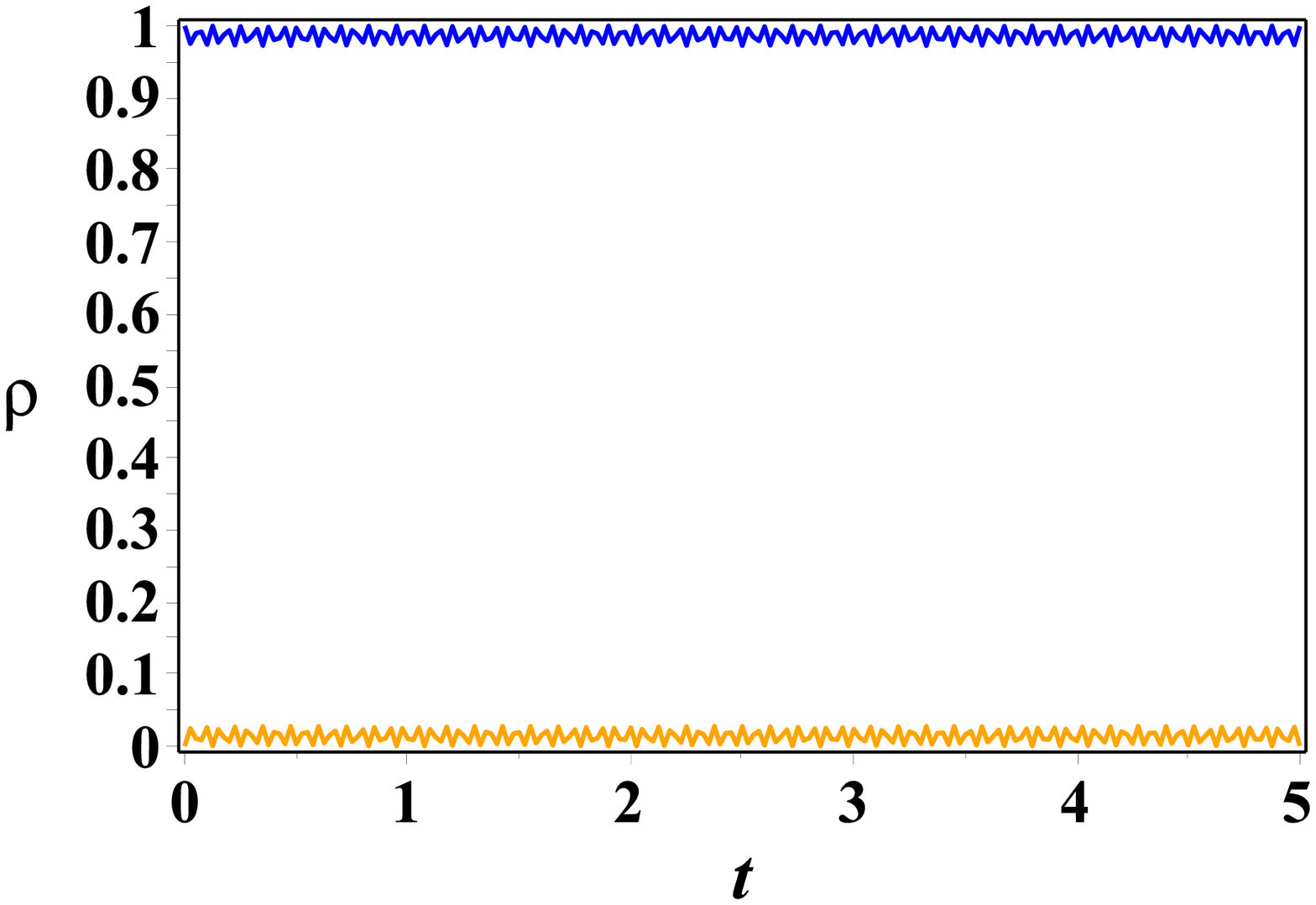}}
 		(a)
 		\scalebox{0.295}{\includegraphics{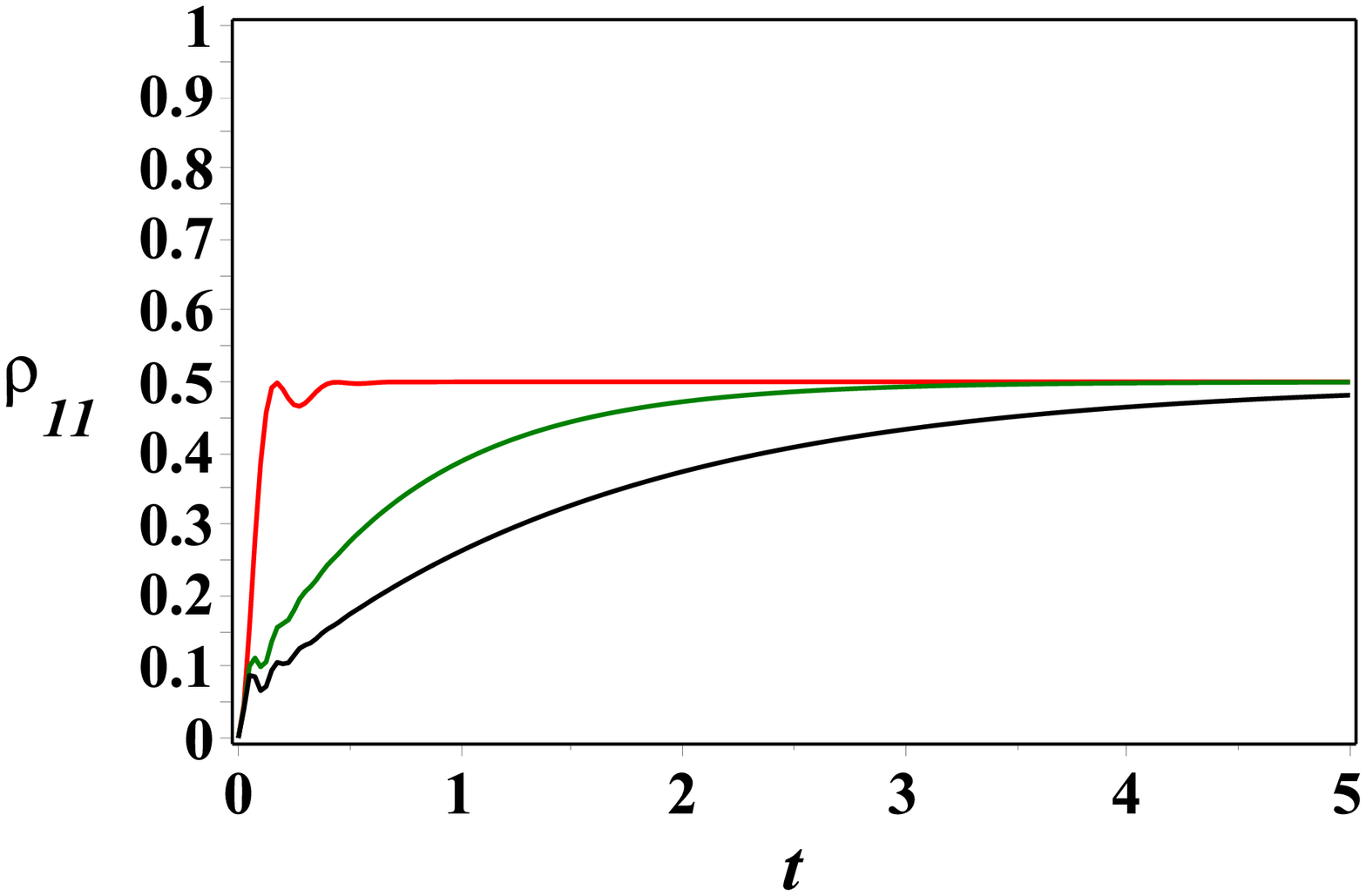}}
 		(b)
 	\end{center}
 	\caption{(Color online) Weakly coupled LHC dimer. Time dependence (in ps) of the density matrix components.  Parameters: $V_{12}=12,6$, $\varepsilon=\varepsilon_2 -\varepsilon_1=150$,  $\gamma=10$. (a) Population of the LHC dimer in the absence of noise: $\rho_{22}(t)$ (blue), $\rho_{11}(t)$ (orange). (b) Population, $\rho_{11}(t)$, in the presence of noise: $d_{21}= 100$ (black), $d_{21}= 150$ (red), $d_{21}= 200$ (green). Initial conditions:  $\rho_{11}(0) =0$,  $\rho_{22}(0) =1$.
 		\label{B1}}
 \end{figure}

\begin{figure}[tbh]
	\begin{center}
		\scalebox{0.285}{\includegraphics{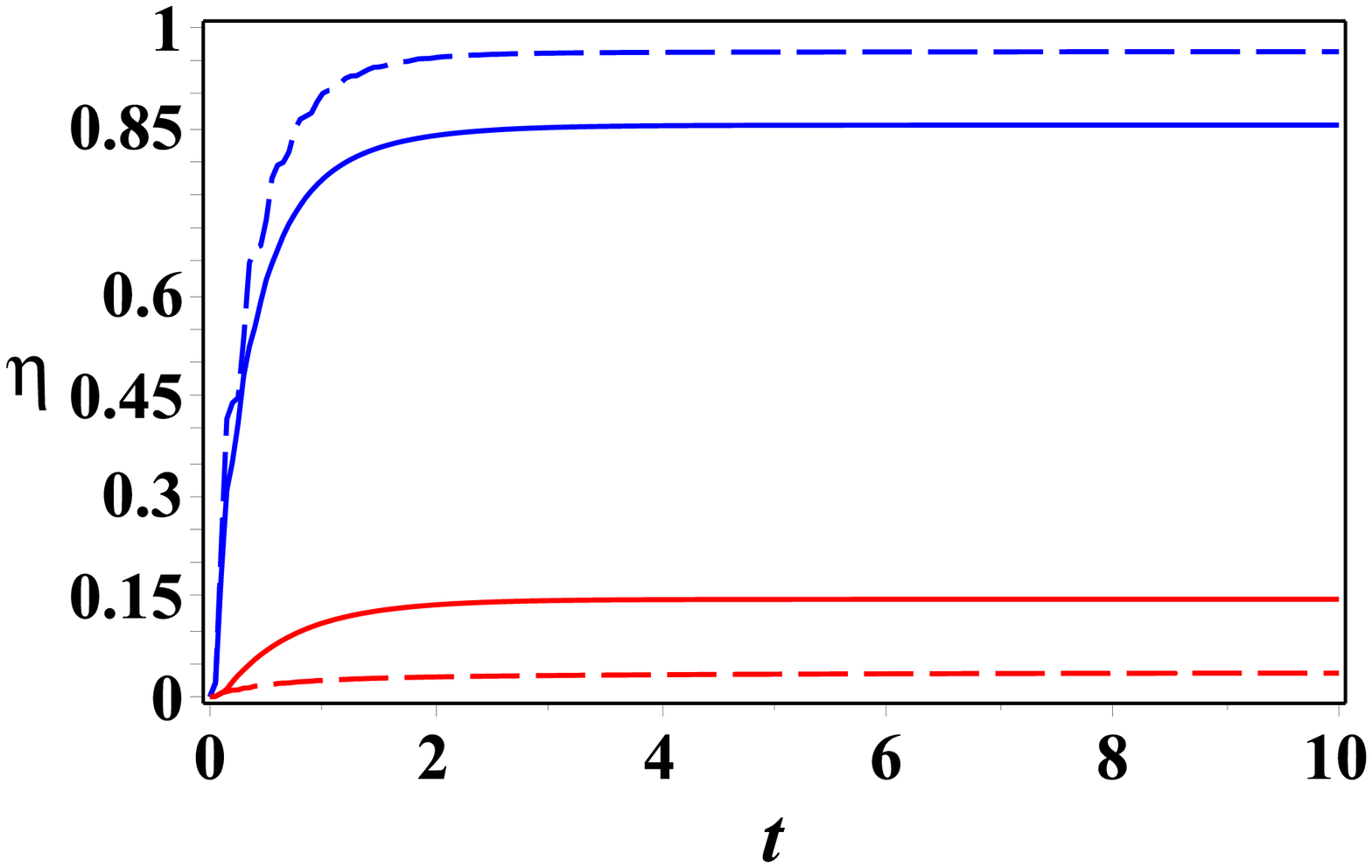}}
		(a)
		\scalebox{0.27}{\includegraphics{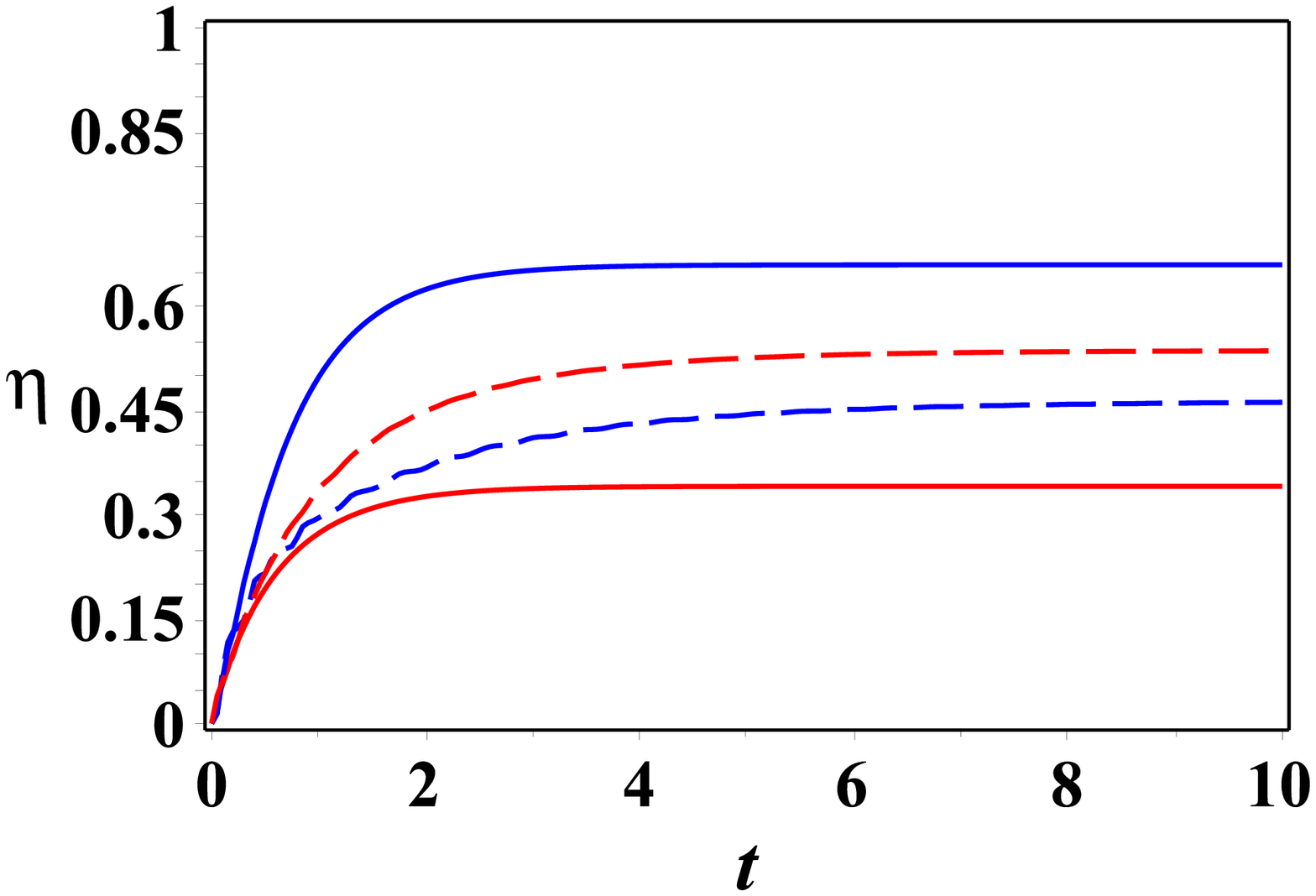}}
		(e)
		\scalebox{0.265}{\includegraphics{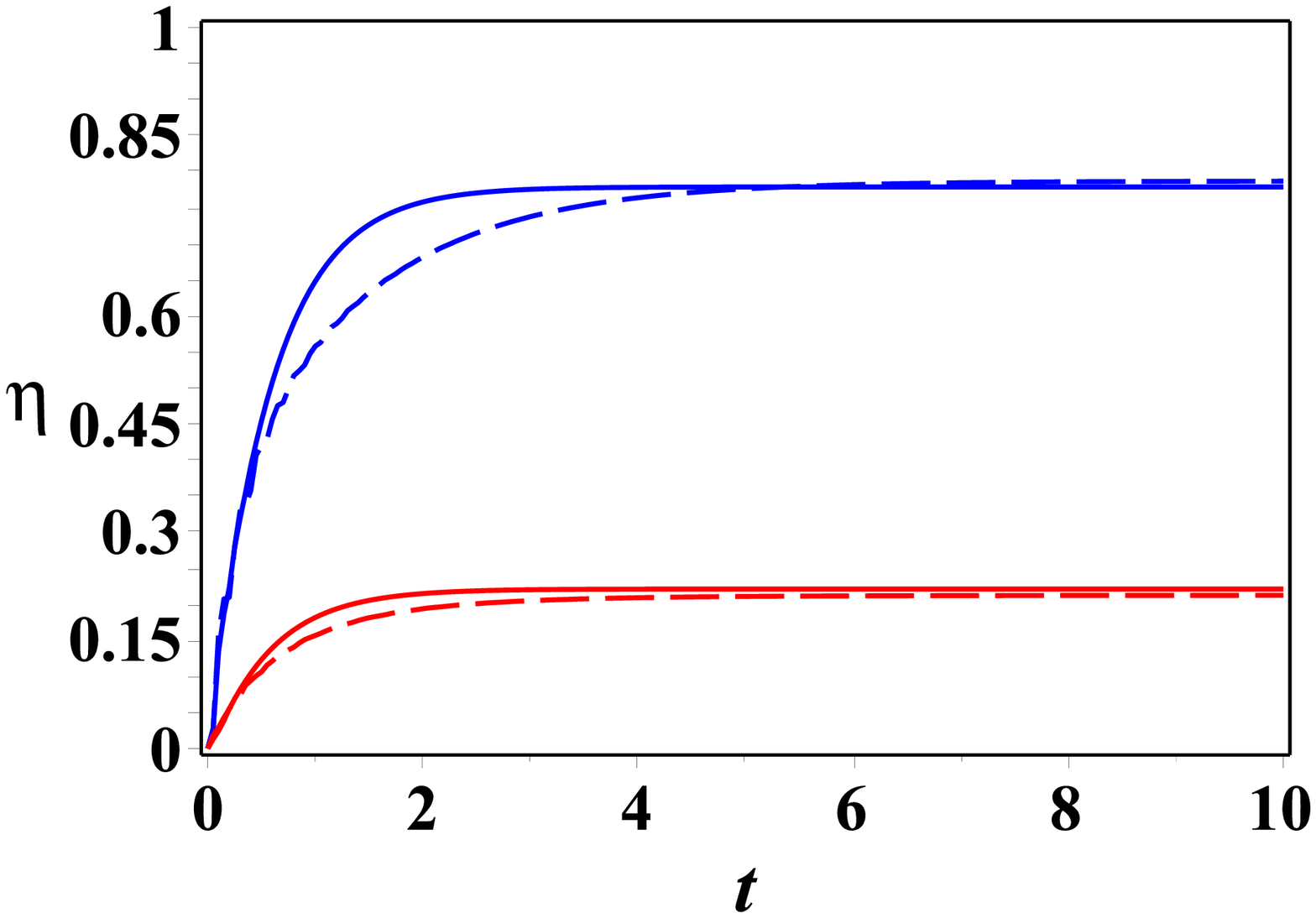}}
		(b)
		\scalebox{0.285}{\includegraphics{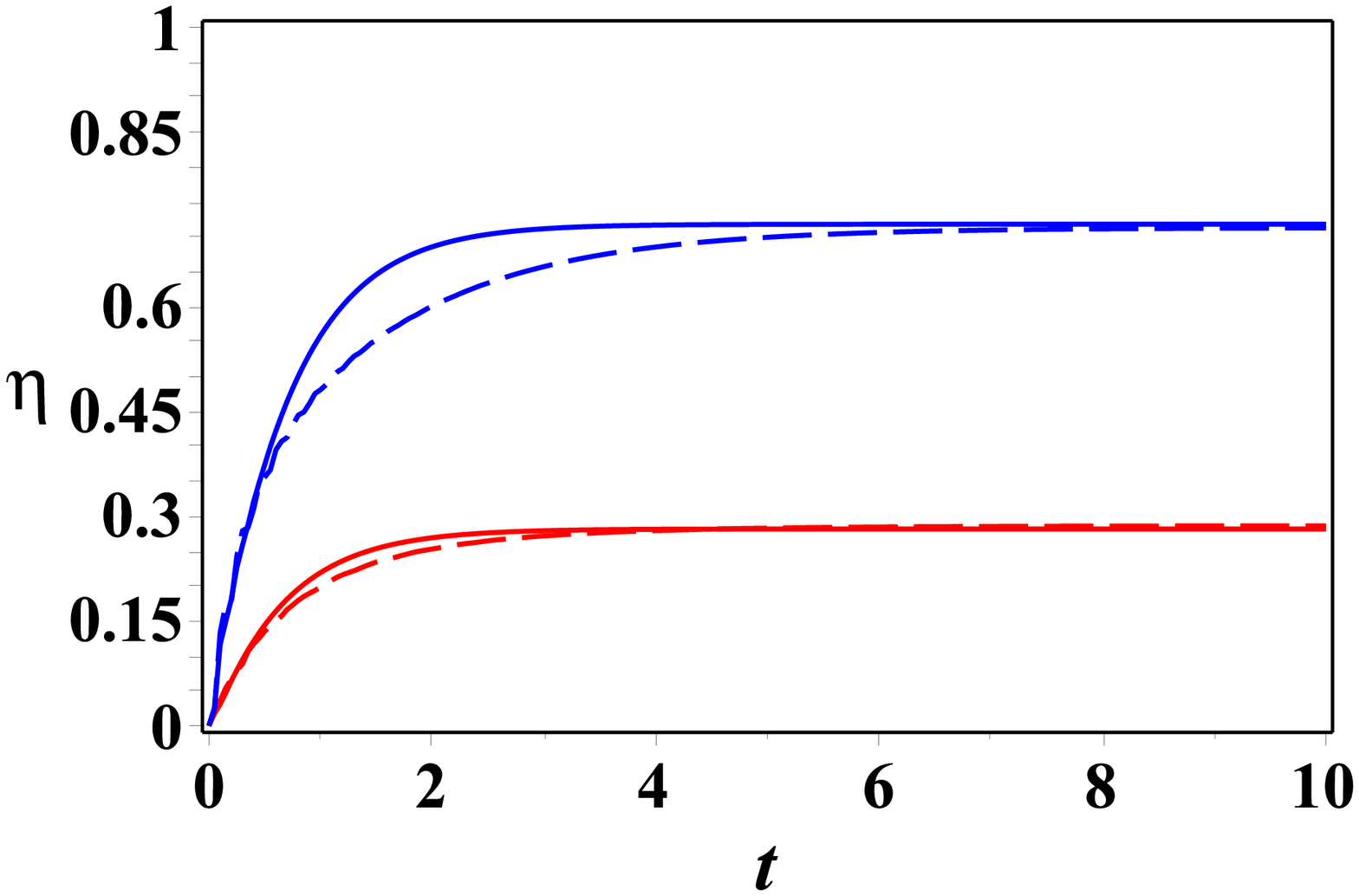}}
		(f)
        \scalebox{0.29}{\includegraphics{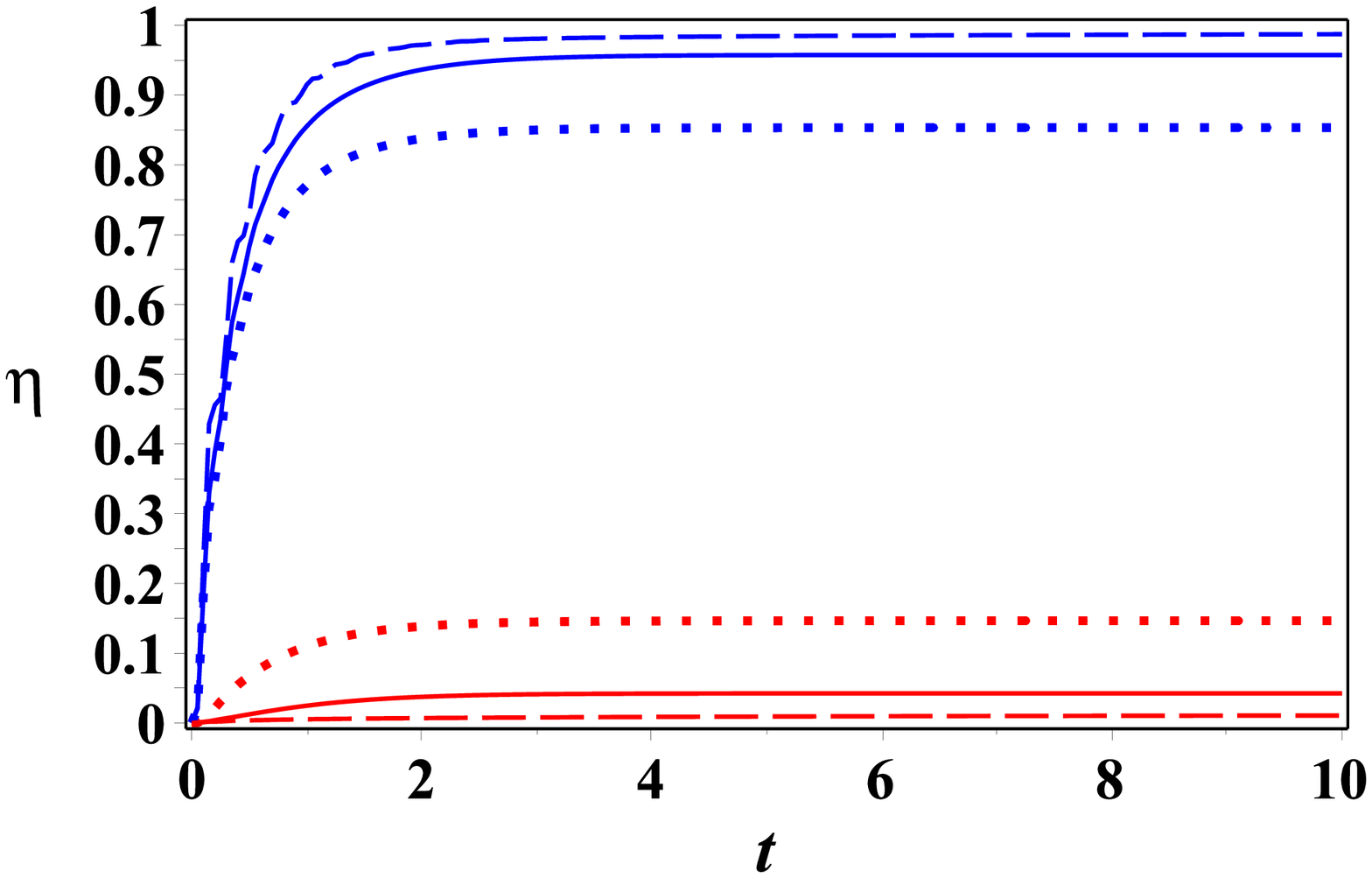}}
		(c)
		 \scalebox{0.28}{\includegraphics{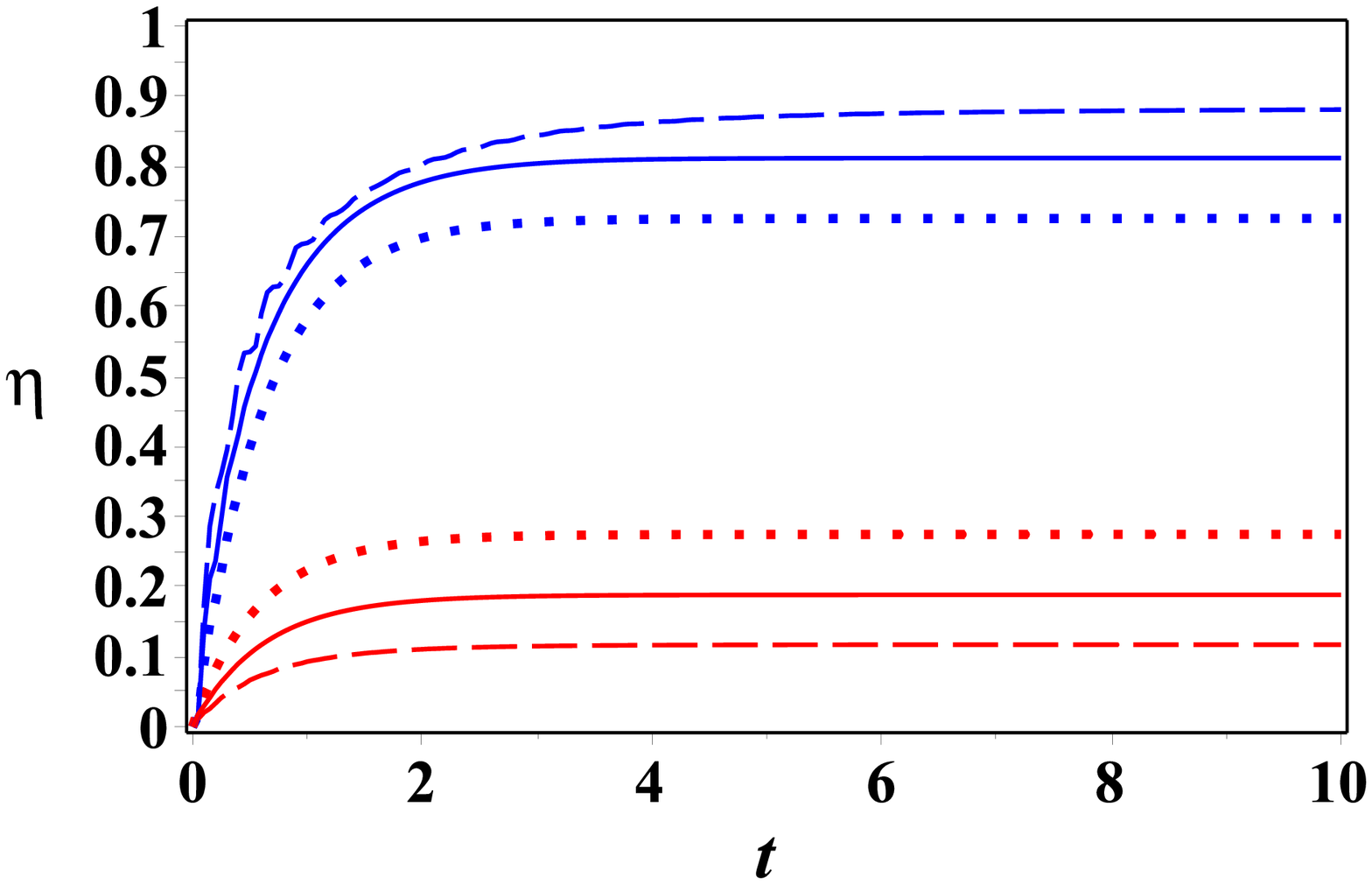}}
		(g)
		\scalebox{0.28}{\includegraphics{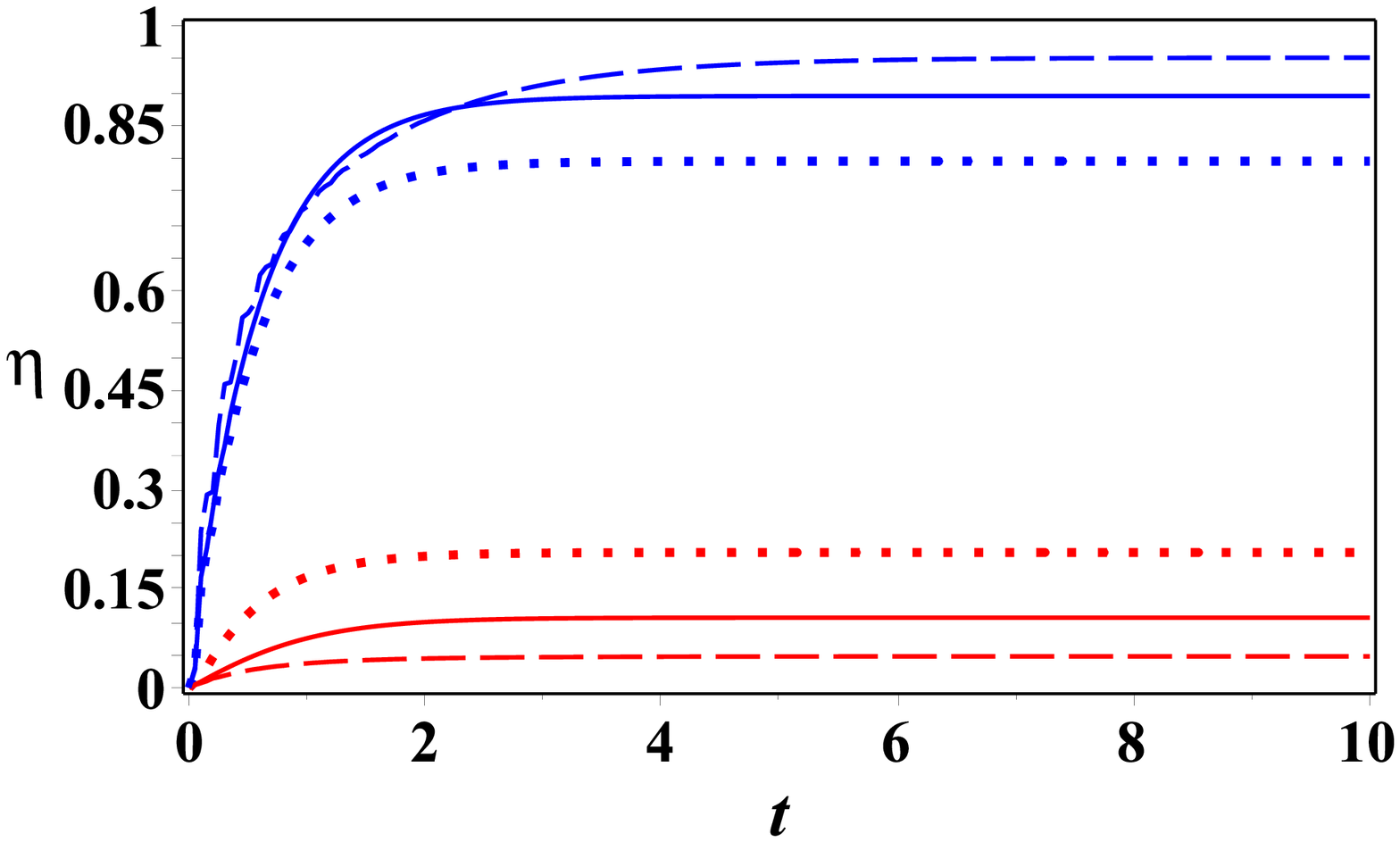}}
		(d)
			\scalebox{0.27}{\includegraphics{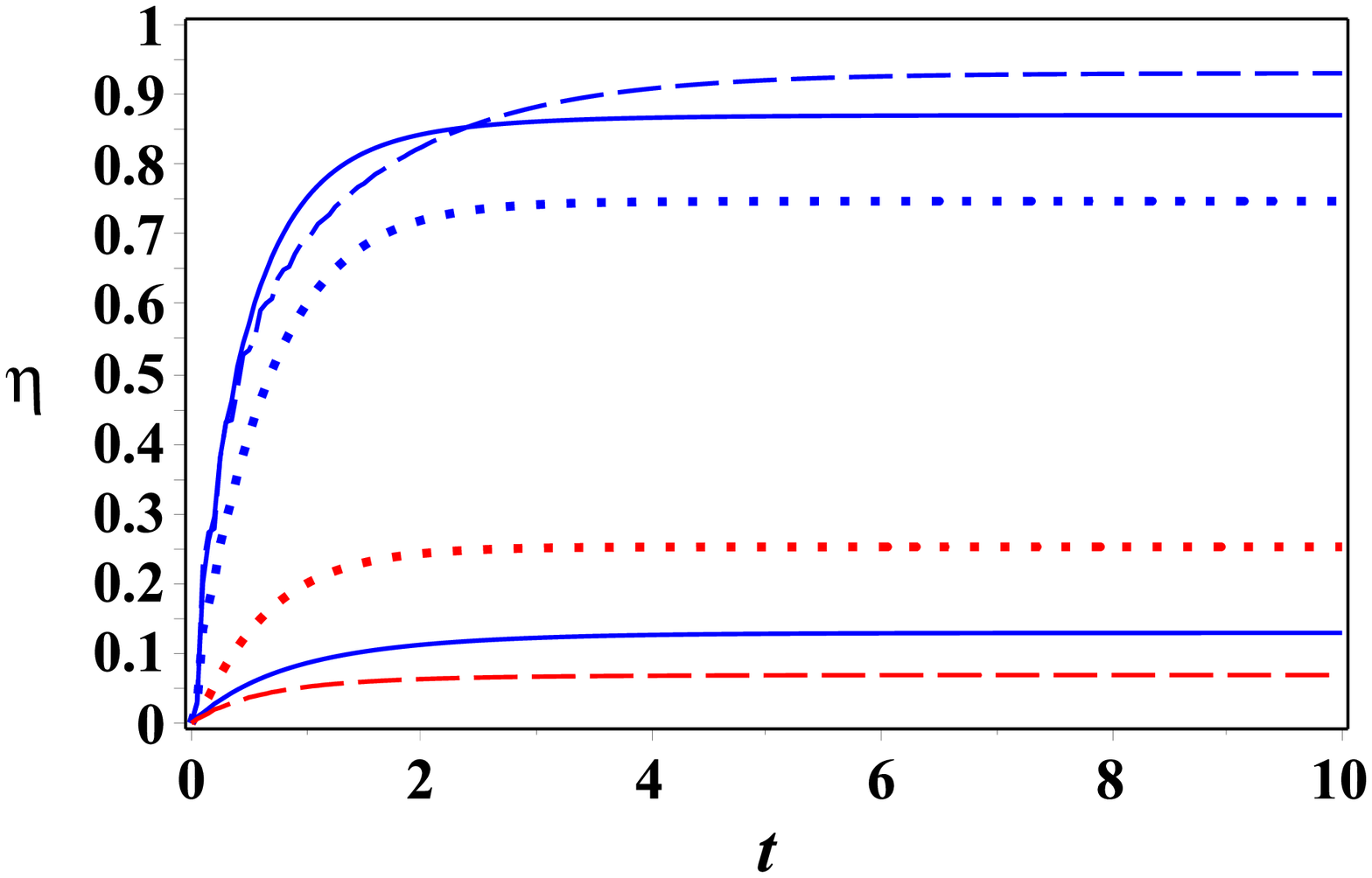}}
		(h)
	\end{center}
	\caption{(Color online) Total system with noise (dashed lines correspond to the system without noise). Time dependences (in ps) of the efficiencies of sinks, $\eta_0(t)$ (red) and $\eta_4(t)$ (blue). Parameters: $V_{10}=V_{20}=30, V_{13} = V_{34}=25,\varepsilon_1=60$, $\varepsilon_3=45$,  $\varepsilon_4=30$, $\Gamma_0 =2$, $\Gamma_4 =10$, $\gamma=10$, $d_0=0, d_1= 60, d_3 =45, d_4 =30$. (a) $V_{12}=20$, $~\varepsilon_0=0$, $\delta = -10$, $d_2 = 70$. (b) $V_{12}=12,6$, $~\varepsilon_0=0$, $\delta = -150$, $d_2 = 210$. (c) $V_{12}=20$, $~\varepsilon_0=-90$, $\delta = -10$, $d_2 = 70$. (d) $V_{12}=12,6$, $~\varepsilon_0=-90$, $\delta = -150$, $d_2 = 210$.  Initial conditions: $|\varphi(0)\rangle=|\varphi_-\rangle$. Left (a,b,c,d): $V_{10}= V_{20}$ . Right (e,f,g,h): $V_{10}= -V_{20}$.  Dotted lines correspond to $d_0=-90$ (resonance noise for the transition $(1-0)$). 
	\label{B6}}
\end{figure}

Assume, that initially the dimer eigenstate, $|\varphi_-\rangle$, is populated. This means that, $c_1(0)=-c_2(0)=1/\sqrt{2}$. Then, $A(0)=0$, and also, $A_0(0)=0$ (as the damaging state, $|0\rangle$, is not initially populated). In this case, $A_0(t)=0$, for all times. This means that due to the ``destructive" quantum interference effects, the damaging state, $|0\rangle$, is not populated (blocked) for all times. Note, that this result does not depend on the difference between the energy of the dimer, $E$, and the energy of the damaging state, $E_0$.

\subsection{Numerical simulations for strongly and weakly coupled dimers}

Below, in numerical simulations, we choose $\hbar=1$. Then, the values of parameters in energy units are measured in $ps^{-1}$, and $1ps^{-1}\approx 0.66meV$. Time is measured in $ps$.

We consider here two cases, of strongly ($\mu_d\gtrsim 1$) and weakly ($\mu_d\ll 1$) coupled chlorophyll dimer. As mentioned above, the additional dimer, $(1-0)$, will be involved in our analysis, which is characterized by two parameters, $\mu_{10}$ and $\delta_{10}$.

In Fig. \ref{A}, we demonstrate the dynamics of the density matrix components for the system with the Hamiltonian  (\ref{HS}) (without a sink and noise), for both strongly and weakly coupled dimers.\\
 {\it Strongly coupled LHC dimer, $(ChlA-ChlB)^*$, $\mu_d=2$}:  In Fig. \ref{A}a, the energy level of the damaging state, $|0\rangle$, is relatively close to the energy level of the $ChlA^*$ (state $|1\rangle$): $\delta_{10}\equiv \varepsilon_1-\varepsilon_0=60$. In this case, the parameter, $\mu_{10}\equiv |V_{10}/\delta_{10}|=0.5$, and the dimer, $(1-0)$, is also strongly coupled. As one can see from Fig. \ref{A}a, when  $V_{10}=V_{20}$, the ``destructive"  interference effects significantly suppress the damaging channel, $|0\rangle$. However, the situation changes completely when $V_{10}=-V_{20}$. (See Fig. \ref{A}e.) This modification  in the population, $\rho_{00}(t)$, of the damaging channel was expected in the considerations in sub-section 6.1.\\
 In Fig. \ref{A}c, the energy level of the damaging state, $|0\rangle$, is located relatively far from  the energy level of the $ChlA^*$ (state $|1\rangle$): $\delta_{10}\equiv \varepsilon_1-\varepsilon_0=150$. In this case, the parameter, $\mu_{10}\equiv |V_{10}/\delta_{10}|=0.084\ll 1$, and the LHC dimer, $(1-0)$, is weakly coupled. Still, as one can see by comparing two dependences, $\rho_{00}(t)$, corresponding to red curves in Fig. \ref{A}c ($V_{10}=V_{20}$) and in Fig. \ref{A}g ($V_{10}=-V_{20}$), the ``destructive" interference effects still contribute significantly to the suppression of the damaging channel.\\
  {\it Weakly coupled LHC dimer, $(ChlA-ChlB)^*$, $\mu_d=0.084\ll 1$}: In Fig. \ref{A}b, the energy level of the damaging state, $|0\rangle$, is located relatively close to the energy level of the $ChlA^*$ (state $|1\rangle$): $\delta_{10}\equiv \varepsilon_1-\varepsilon_0=60$. In this case, the parameter, $\mu_{10}\equiv |V_{10}/\delta_{10}|=0.5$, and the LHC dimer, $(1-0)$, is strongly coupled. The initial condition, $|\varphi_-\rangle$, corresponds to the initial population of the state, $|1\rangle$, with high probability, $\rho_{11}(0)=0.993$. Then, the dynamics  approximately follows that of the two-level system, $(1-0)$.  Using the well-know expression for the two-level system, $\rho^{max}_{00}\approx 4V_{10}^2/(\delta^2_{10}+4V_{10}^2)\times\rho_{11}(0)$, we have: $\rho_{00}^{max} \approx 0.5$. By comparing Fig. \ref{A}b and Fig. \ref{A}f, one can see that the interference effects do not play a role in the dynamics of the damaging channel (red curves).\\
Finally, we consider the case of a weakly coupled dimer, presented in Fig. \ref{A}d.   The energy level of the damaging state, $|0\rangle$, is located relatively far from  the energy level of the $ChlA^*$ (state $|1\rangle$): $\delta_{10}\equiv \varepsilon_1-\varepsilon_0=150$. (Similar to the case presented in Fig. \ref{A}c.) In this case, the parameter, $\mu_{10}\equiv |V_{10}/\delta_{10}|=0.084\ll 1$, and the LHC dimer, $(1-0)$, is weakly coupled. The dynamics follows approximately that of the effective two-level systems, $|1\rangle-|0\rangle$. So, we have: $\rho^{max}_{00}\approx 4V_{10}^2/(\delta^2_{10}+4V_{10}^2)\times\rho_{11}(0)\approx 0.14$.  As one can see by comparison two dependences, $\rho_{00}(t)$, corresponding to the red curves in Fig. \ref{A}d ($V_{10}=V_{20}$) and in Fig. \ref{A}h ($V_{10}=-V_{20}$), the ``destructive" interference effects give a very little contribution to the suppression of the damaging channel.

One conclusion from the above results could be the following. For a strongly coupled dimer, quantum interference effects could significantly contribute to the suppression of the damaging channel. For a weakly coupled dimer, quantum interference effects do not practically contribute to the suppression of the damaging channel.  Below, these conclusions will be confirmed when considering the total system, presented in Fig. \ref{S2}. 
  
 \subsection{Non-NPQ regime: Influence of sink} 
 In this sub-section, we add the sink, $|S_0\rangle$, to the three-level system shown in Fig. \ref{S2a}. The sink is characterizes by two parameters (see Sec. 3), the rate, $\Gamma_0$, and the efficiency, $\eta_0(t)$, 
 \begin{equation}
 \eta_0(t)=\Gamma_0\int_0^t\rho_{00}(\tau)d\tau.
\label{eta}
 \end{equation}  
 In Fig. \ref{A4}, we demonstrate the results of the  numerical simulation of the system presented in Fig. \ref{S2a}, with the same parameters and initial conditions  used in Fig. \ref{A}a-d. As was discussed in  sub-section 6.2, this corresponds to both strongly and weakly coupled dimer, and with the additional condition, $V_{10}=V_{20}$. This last condition results in the influence of the interference effects on suppression of the damaging channel, $|0\rangle$ (without sink) for a strongly coupled dimer.

 As one can see from Eq. (\ref{eta}), the efficiency of the sink, $|S_0\rangle$, on the dynamics of the system depends significantly on the population, $\rho_{00}(t)$, of the discrete damaging level, $|0\rangle$. Then, we should expect that in the  two cases of strongly coupled dimer, shown in Figs. \ref{A}a,c,  the efficiency, $\eta_0(t)$, of the damaging channel will be less than in all other cases. Our simulations presented in Fig. \ref{A4}  demonstrate this is indeed the case. We have chosen the rate, $\Gamma_0=2$. For parameters chosen in Fig. \ref{A4}a,  the  saturation of the damaging channel is rather slow, $t_{sat} >500ps$. For the parameters chosen in Fig. \ref{A4}c,  the dynamics of the saturation of the damaging channel is even  slower, $t_{sat} >1000ps$. But, as expected, the saturation dynamics of the damaging channel becomes significantly faster for parameters chosen in Figs. \ref{A4}b,d, $t_{sat}\approx 20ps$ and $t_{sat}\approx 100ps$, correspondingly.
 
\subsection{Influence of noise on the dimer population}
In this sub-section we briefly discuss the dynamics of population of the isolated LHC chlorophyll dimer, $(ChlA-ChlB)^*$,  under the influence of noise. So, the discrete level, $|0\rangle$, and the sink, $|S_0\rangle$, are absent.

Suppose that the LHC dimer, presented in Fig. \ref{S2a}, has the parameters:  $\varepsilon\equiv\varepsilon_2-\varepsilon_1=150ps^{-1}\approx 99meV$ and $V_{12}=12.6ps^{-1}\approx 8.3meV$. This is the case for the LHC dimer based on $Chlb606$ and $Chla604$ in the CP29 \cite{Muh}. Then, the parameter, $\mu_d=0.084\ll 1$, and the dimer is weakly coupled. Suppose that initially the $Chlb606$ is populated, $\rho_{22}(0)=1$ and $\rho_{11}(0)=0$.  Without the influence of the environment (noise in our case), the population of the $Chla604$ will remain very small for all times, as shown in Fig. \ref{B1}a. As it is well-known, the influence of noise significantly changes this result. (See Fig. \ref{B1}b.) The important issue is the optimal amplitude of noise which provides a maximal rate of approach the stationary state. As one can see from  Fig. \ref{B1}b, this ``resonant" amplitude of noise is: $d_{21}\approx\varepsilon$, where $d_{21}\equiv(d_2-d_1)\sigma$. In this case, the stationary regime (saturation) is approached at times,  $t_{sat}\approx 0.5ps$. (See also the experimental results \cite{Eads}.) Qualitatively, a similar ``resonant" influence of the thermal environment is contained in the well-known Marcus theory for ET rates: $\varepsilon\approx\lambda$, where $\lambda$ is the reconstruction energy. (See, for example, \cite{Xu}, and references therein.)

\section{Results of numerical simulations for the total system}
\begin{figure}[tbh]
	\begin{center}
		\scalebox{0.285}{\includegraphics{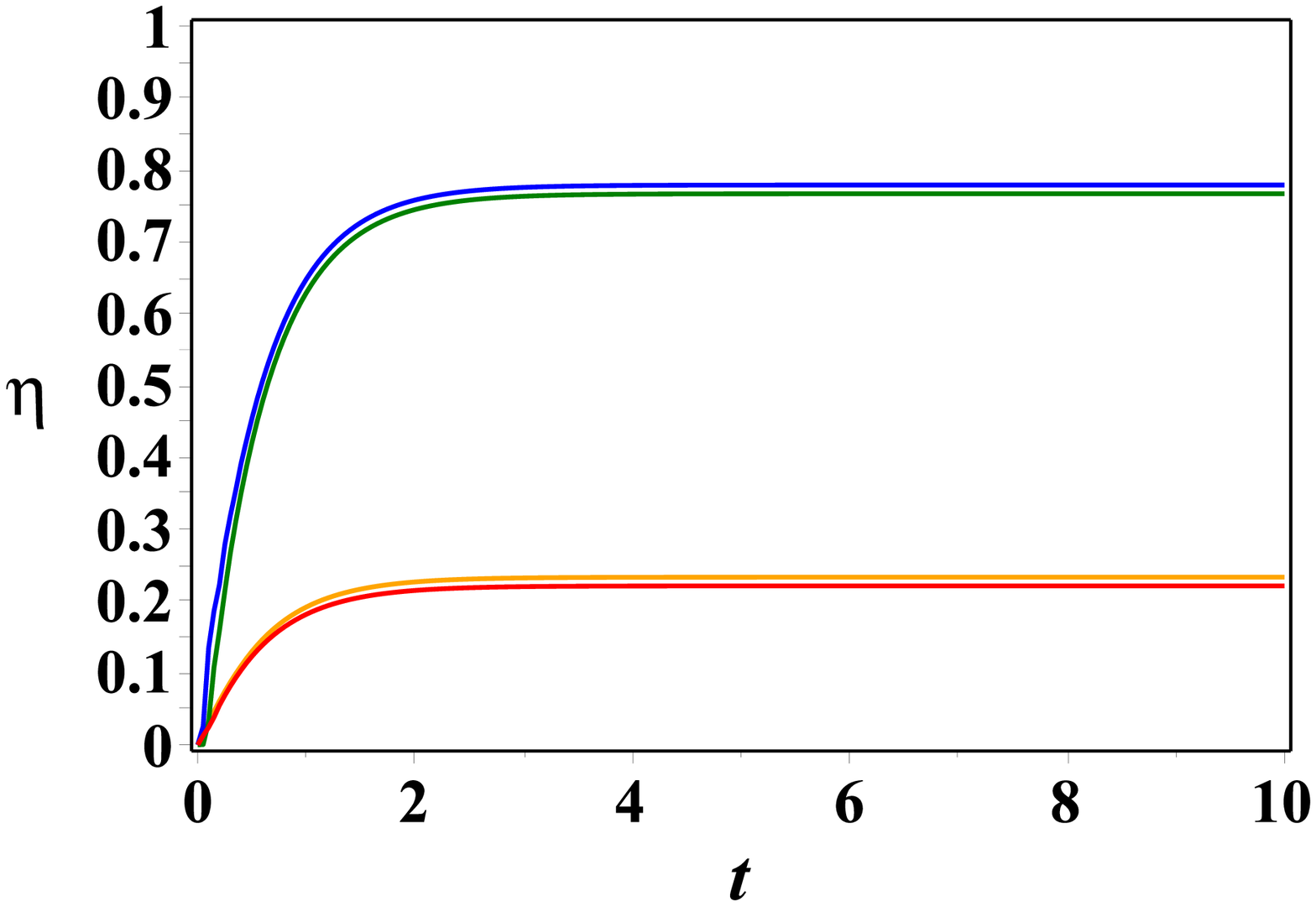}}
		(a)
		\scalebox{0.29}{\includegraphics{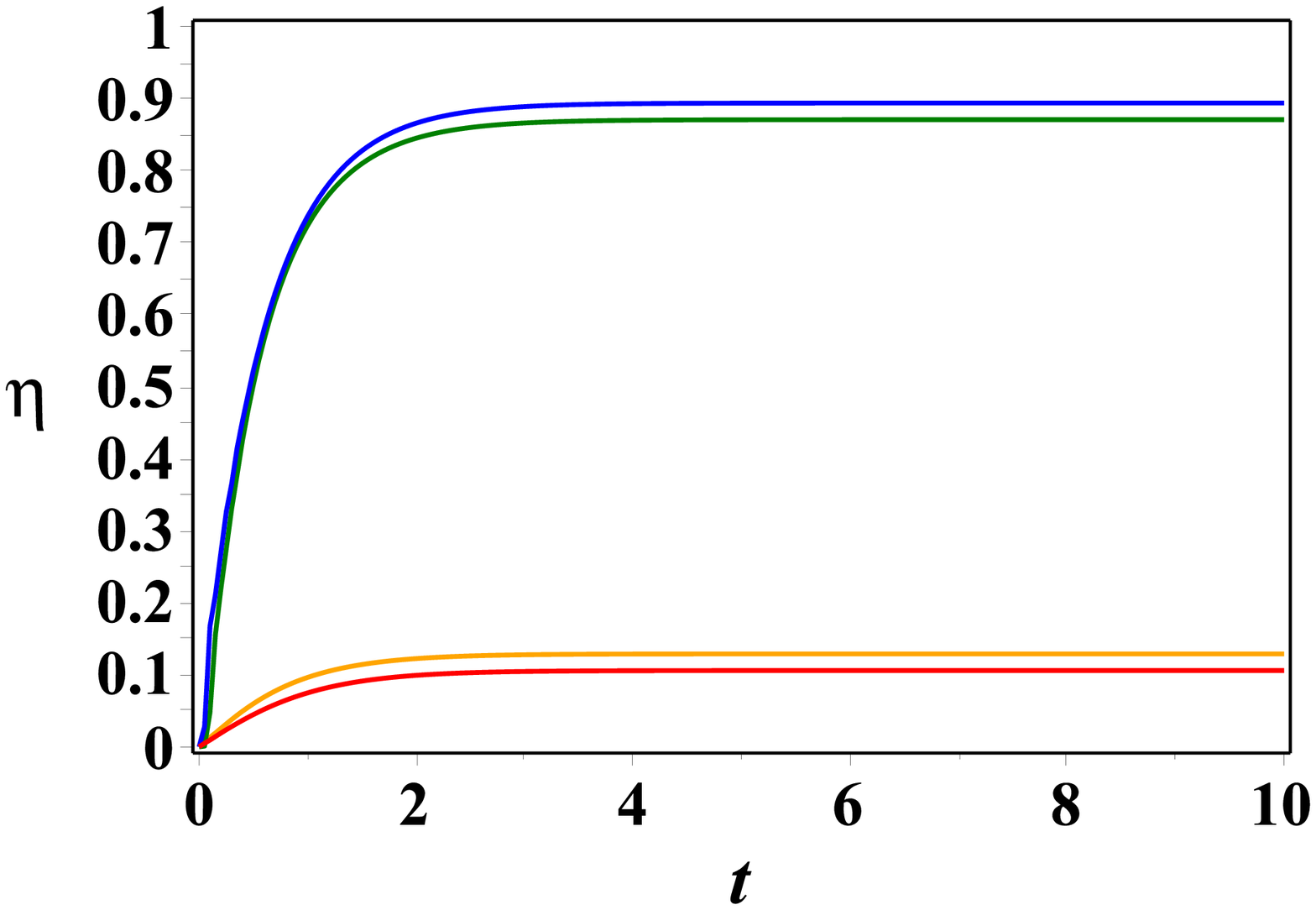}}
		(b)
	\end{center}
	\caption{(Color online) Total system with noise. Time dependencies (in ps) of the efficiencies of sinks, $\eta_0(t)$ (red, orange) and $\eta_4(t)$ (blue, green). Parameters: $V_{10}=V_{20}=30, V_{13} = V_{34}=25,\varepsilon_1=60$, $\varepsilon_3=45$,  $\varepsilon_4=30$, $\Gamma_0 =2$, $\Gamma_4 =10$, $d_0=0, d_1= 60, d_3 =45, d_4 =30$.  (a) $V_{12}=12,6$, $~\varepsilon_0=0$, $\delta = -150$, $d_2 = 210$.  (b) $V_{12}=12,6$, $~\varepsilon_0=-90$, $\delta = -150$, $d_2 = 210$.  Initial conditions: $|\varphi(0)\rangle=|\varphi_-\rangle$ (blue, red). Initial conditions: $|\varphi(0)\rangle=|\varphi_{+}\rangle$ (green, orange).
		\label{B7}}
\end{figure}

\begin{figure}[tbh]
	\begin{center}
		\scalebox{0.285}{\includegraphics{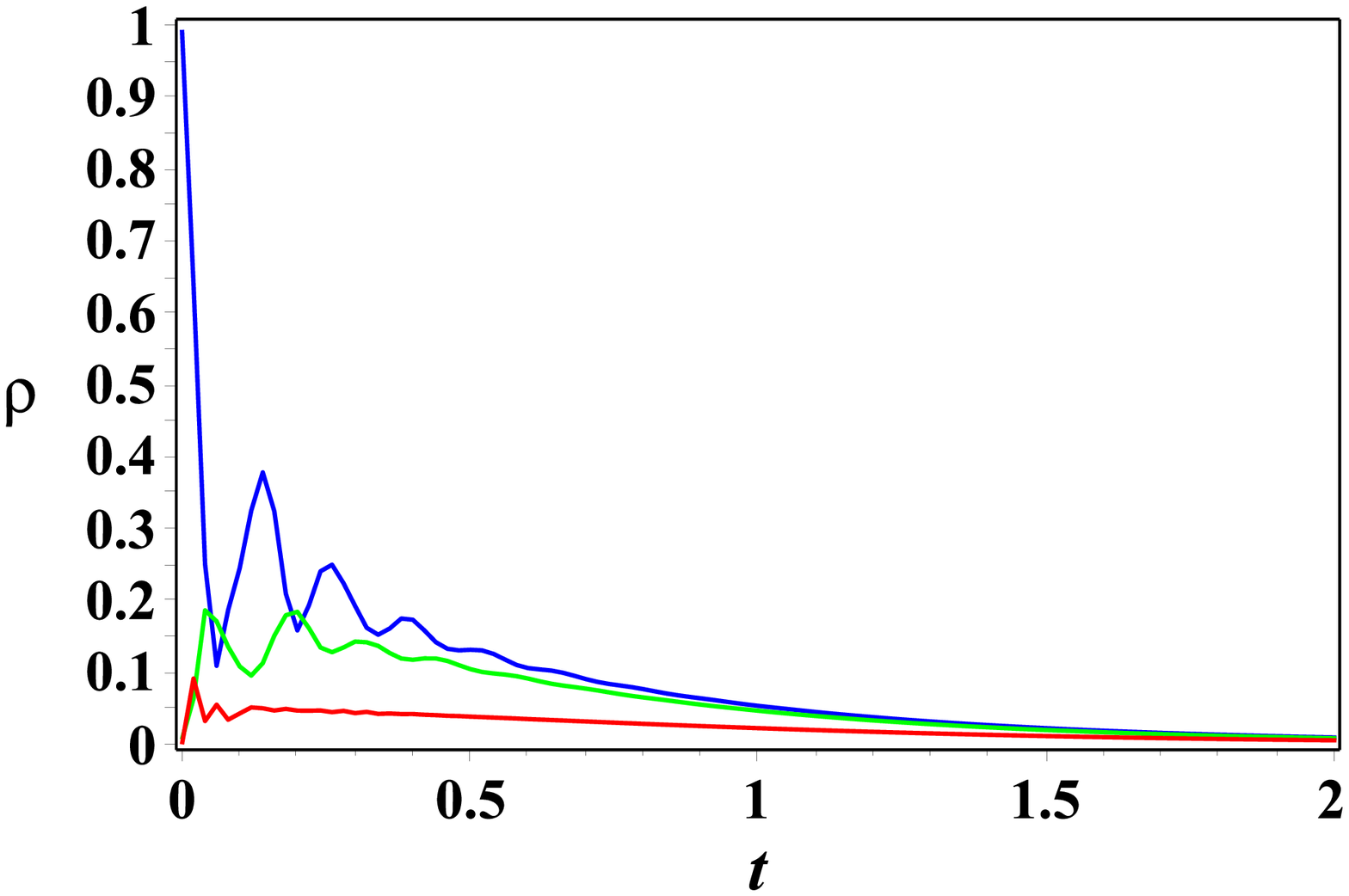}}
		(a)
		\scalebox{0.29}{\includegraphics{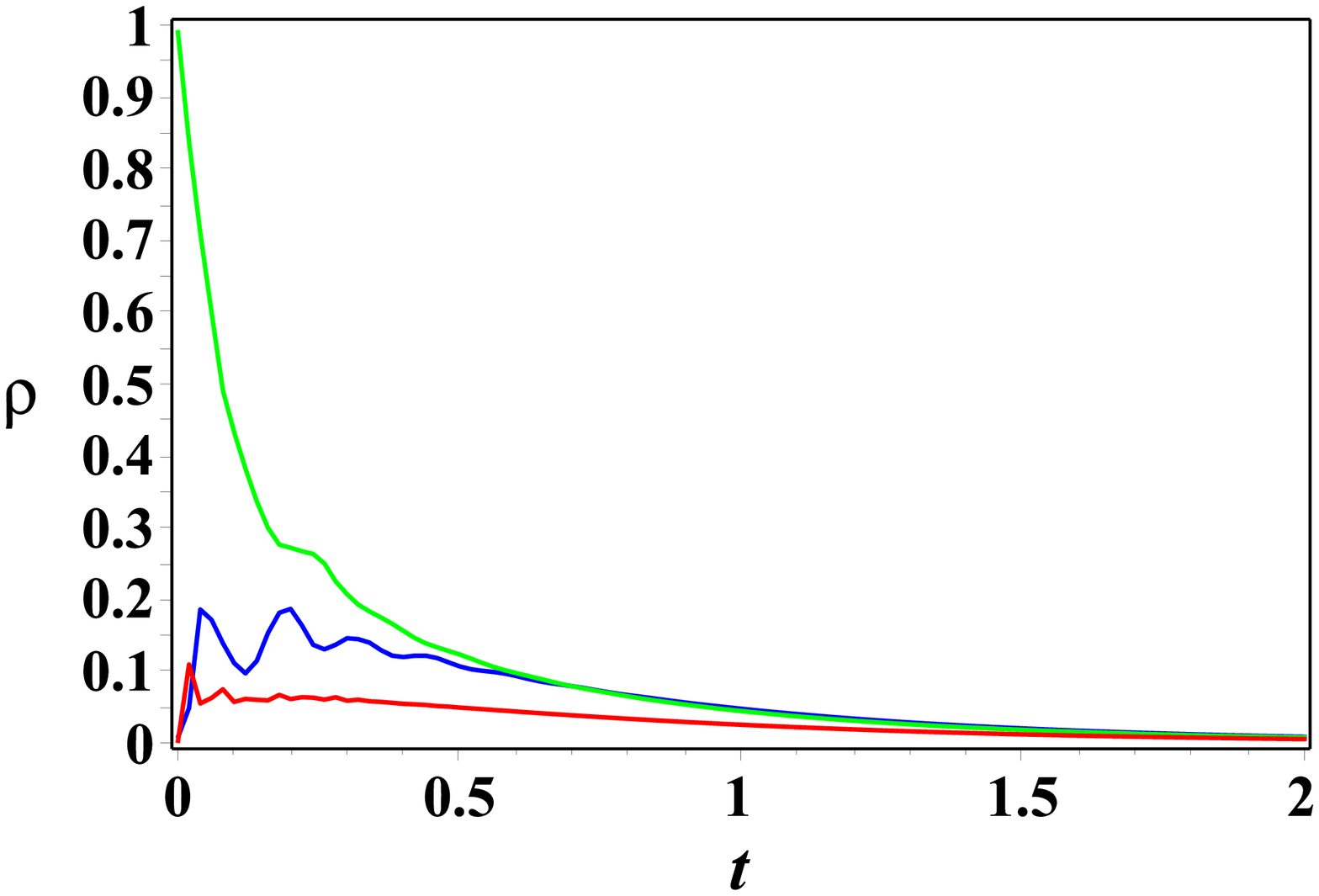}}
		(b)
	\end{center}
	\caption{(Color online) Total system with noise. Time dependence (in ps) of the averaged density matrix components: $\langle \rho_{00}(t)\rangle $ (red), $\langle\rho_{11}(t) \rangle$ (blue), $\langle\rho_{22}(t) \rangle$ (green). (a) Initial conditions:  $|\varphi(0)\rangle=|\varphi_-\rangle$. (b) Initial conditions: $|\varphi(0)\rangle=|\varphi_{+}\rangle$ . Parameters: $V_{10}=V_{20}=30$, $V_{12}=12,6$, $V_{13} = V_{34}=25$, $\varepsilon_0=-90$, $\varepsilon_1=60$, $\varepsilon_2=210$, $\varepsilon_3=45$, $\varepsilon_4=30$, $\Gamma_0 =2$, $\Gamma_4 =10$, $d_0=0$, $d_1= 60$, $d_2 = 210$, $d_3 =45$, $d_4 =30$.  		
		\label{B8}}
\end{figure}

 Here we present the results of our numerical simulations for the complete model shown in Fig. \ref{S2}, which includes five discrete energy levels, two sinks, and noise. Our simulations were done by numerically solving the equations, (\ref{IB4}) and (\ref{IB5}), for the average components of the density matrix, $\langle\rho_{mn}(t)\rangle$ $(n,m=0,...,4$), and for the efficiencies of the sinks, $\eta_{0,4}(t)$. 
 
 As follows from Eqs. (\ref{ET1}) and (\ref{Eq16ar}), for the complete system with two sinks and noise, the following normalization condition is satisfied,
 \begin{equation}
 	\label{C1}
 	\sum_{n=0}^4 \langle\rho_{nn}(t)\rangle+\eta_0(t)+\eta_4(t)=1,
 \end{equation}
 where the efficiencies of the sinks, $|S_0\rangle$ and $|S_4\rangle$, are,
 \begin{equation}
 	\label{Ef}
 	\eta_0(t)= \Gamma_0\int_0^t \langle\rho_{00}(\tau) \rangle d\tau,~\eta_4(t)= \Gamma_4\int_0^t \langle\rho_{44}(\tau) \rangle d\tau,
 \end{equation}
 and $\Gamma_0 $ and $\Gamma_4$, are the rates into the corresponding sinks.
 The normalization condition, (\ref{C1}),  requires that the total probability of finding the electron among five discrete levels and in two sinks is unity for all times. 
 
 In order to implement successfully the NPQ mechanism in our model, based on the interference  and sink effects discussed above, some  conditions must be satisfied. First, the implementation of the NPQ mechanism requires  small values for both, $\langle\rho_{00}(t)\rangle$ (probability on the damaging discrete state, $|0\rangle$) and for  $\eta_0(t)$ (cumulative probability on the damaging sink, $|S_0\rangle$). In our model, the quenching due to interference and sinks means that the initial energy accumulated in the  state of the dimer (say, $|\varphi_-\rangle$) does not transfer efficiently into the damaging state, $|0\rangle$, and its sink, $|S_0\rangle$, but instead is efficiently transfered to the discrete states, $|3\rangle$ (heterodimer) and $|4\rangle$ (charge state),  and to the sink, $|S_4\rangle$.

In Fig. \ref{B6}, we demonstrate the typical results of numerical simulations for the efficiencies of both sinks, $\eta_0(t)$ (red curves) and $\eta_4(t)$ (blue curves), for the total system shown in Fig. \ref{S2}, with two sinks and noise. (Dashed curves correspond to the absence of noise.) Both strongly and weakly coupled LHC dimer were simulated. The parameters of the dimers and the damaging state, $|0\rangle$, are the same as in Fig. \ref{A}. Additionally, the discrete states, $|3\rangle$ and $|4\rangle$, were added, and their sinks, $|S_0\rangle$ and $|S_4\rangle$, correspondingly. The amplitudes of noise, $d_n$ $(n=0,...,4$), were chosen close to those discussed in the literature. In particular, the discussed in \cite{Ber6} ``resonant conditions" $(d_2-d_1=\varepsilon_2-\varepsilon_1$) were used for the dimer transition, $|2\rangle\longleftrightarrow|1\rangle$, for both strongly and weakly coupled LHC dimers. This provides a rapid re-population of the states, $1\rangle$ and $2\rangle$. (See Fig. \ref{B1}b.) The noise correlation decay rate was chosen to be, $\gamma=10ps^{-1}$. The rates to sinks were: $\Gamma_0=2ps^{-1}$ and $\Gamma_4=10ps^{-1}$. The left panel in Fig. \ref{B6}, corresponds to the choice of matrix elements, $V_{10}=V_{20}$, for which  the interference effects reveal themselves for a strongly coupled dimer. The right panel corresponds to the choice of matrix elements, $V_{10}=-V_{20}$, for which the interference effects do not influence suppression of the damaging channel. 

As one can see from the results presented in Fig. \ref{B6}, the saturation of both efficiencies, $\eta_0(t)$ and $\eta_4(t)$, appears at relatively short times, $t_{sat}=2-4ps$. Note that in the saturation regime, all density matrix elements approach zero.  For a strongly coupled LHC dimer (Figs. \ref{B6}a,c), the interference effects contribute to a suppression of the damaging channel, $|S_0\rangle$. (Compare Figs. \ref{A6}a,c and Figs. \ref{B6}e,g.) The best NPQ regime corresponds to a strongly coupled LHC dimer, with the energy of the damaging state far from the energy lever of the $ChlA^*$ (Fig. \ref{B6}c). But also for a weakly coupled LHC dimer, with the energy of the damaging state far from the energy lever of the $ChlA^*$, the contribution of the damaging channel, $|S_0\rangle$, in the NPQ process is relatively small, $\eta_0(t)\approx 10\%$.

In Figs. \ref{B6}c,d,g,h (dotted lines), we present the results for the ``resonant" condition,
$d_1-d_0=\varepsilon_1-\varepsilon_0$. In this case, the population, $\rho_{00}(t)$, of the damaging state increases, resulting in increasing the efficiency, $\eta_0$, of the damaging channel. As the result, NPQ performance drops to $\eta_4\approx 80\%$. 

Concluding this section, we note that, according to our results and for chosen parameters, the efficiency of the NPQ mechanism by a charge transfer state can achieve, $\eta_4\approx 80-90\%$.

\subsection{Dependence on the initial population of the weakly coupled LHC dimer}
	As demonstrated above for a weakly coupled LHC dimer, the interference effects do not play a role in the NPQ regime. (Compare Figs. \ref{A}b,d and Figs. \ref{A}f,h.) The question remains of whether or not the initial population of either the  $ChlA^*$ or $ChlB^*$ of the weakly coupled LHC dimer will make a significant difference in the performance of the NPQ mechanism. 	The results of the numerical simulations on the efficiencies of both sinks are shown in Fig. \ref{B7}. We have chosen a weakly coupled LHC dimer, with $\delta=\varepsilon_1-\varepsilon_2=-150ps^{-1}\approx -99meV$.  In Fig. \ref{B7}a, the damaging state, $|0\rangle$, was below the state, $|1\rangle$, by $60ps^{-1}\approx 39.6meV$. In Fig. \ref{B7}b, the damaging state, $|0\rangle$, was below the state, $|1\rangle$, by $150ps^{-1}\approx 99meV$. The blue and red curves in Fig. \ref{B7} describe the dynamics of $\eta_0(t)$ and $\eta_4(t)$, correspondingly. These curves correspond to the initial population of the LHC dimer eigenstate, $|\varphi_-\rangle$. (See Figs. \ref{A},b,d.) This initial condition corresponds to the initial population of the $ChlA^*$ with high probability, $\rho_{11}\approx 0.993$. The green and orange curves in Fig. \ref{B7} correspond to the initial population of the dimer eigenstate, $|\varphi_+\rangle$.  This initial condition corresponds to the initial population of the $ChlB^*$ with the probability, $\rho_{22}\approx 0.993$. As one can see from Fig. \ref{B7}, the dynamics of the efficiencies of both sinks, is essentially independent of  the chosen initial conditions. 
	
	Even though the dynamics of the efficiencies, $\eta_0(t)$ and $\eta_4(t)$, are approximately the same in Fig. \ref{B7}, for both initial population of the weakly coupled LHC dimer, the time-dependences  of the density matrix elements are different. In Fig. \ref{B8}, all parameters were chosen as in Fig. \ref{B7}b (weakly coupled dimer with a low energy level, $|0\rangle$). Fig. \ref{B8}a corresponds to the initial population: $|\varphi(0)\rangle=|\varphi_-\rangle\approx|1\rangle$. Fig. \ref{B8}b corresponds to the initial population: $|\varphi(0)\rangle=|\varphi_+\rangle\approx|2\rangle$. As one can see, the dynamics of the density matrix components, $\rho_{11}(t)$ (blue curves) and $\rho_{22}(t)$ (green curves), are different for different initial conditions. At the same time, the dynamics of the population of the damaging state, $\rho_{00}(t)$ is  similar for both initial conditions,  resulting in similar dependences of the efficiencies shown in Fig. \ref{B7}, for different initial conditions.
	
	We also would like to mention the presence of coherent quantum oscillations in our numerical simulations for both diagonal and non-diagonal density matrix elements, similar to those shown in  Fig. \ref{B8}. These oscillations demonstrate that even in the presence of sinks and noise, the quantum dynamics in the NPQ mode reveals that both damaging and charge transfer processes are partly reversible.

\section{Conclusion}
We introduced a  simple quantum-mechanical model which is intended to better understand a possible role of interference and sink effects in NPQ processes in LHCs. Our model includes two sites of the LHC dimer based on the excitonic states of $ChA^*$, and $ChlB^*$, the sites of the heterodimer, $(ChlA-Zea)^*$, and its charge transfer state $(ChlA^--Zea^+)^*$, and two sinks. One sink is responsible for the damaging channel, and the second sink is responsible for the charge transfer processes. The protein environment is modeled by a random telegraph process. The implementation of the NPQ regime requires that the efficiency of the charge transfer sink significantly exceeds the efficiency of the damaging sink. Our numerical simulations demonstrate that the efficiency of the charge transfer channel can approach $80-90\%$, for reasonable parameters of the system. Still, the efficiency of the NPQ regime can be improved by optimizing the parameters of the system. We also demonstrated that if the dimer is strongly coupled,   interference effects can play a significant role in the performance of the NPQ mechanism. However, for a weakly coupled dimer, the interference effects are not significant. A noisy environment plays a significant role in the NPQ mode, as it provides an effective  re-population of the exciton among the chlorophylls, the heterodimer states, and sinks.  

To verify how the effects of the interference and sinks  are related to the quenching mechanisms in natural light-harvesting bio-complexes, additional experimental studies are needed.  The main question is whether or not the formation of the LHC dimer based on the excited states of two chlorophylls indeed is important in the NPQ mode. (See discussions on this issue in \cite{Ahn,Duffy2}, and in references therein.) The LHC dimer, $(ChlA_5-ChlB_5)^*$, in the CP29 minor complex, is characterized by two parameters: (i) the energy gap, $\delta$, between the excited states of two non-interacting chlorophylls, and (ii) the matrix element, $V_{AB}$, of their interaction. If the dimensionless parameter, $|V_{AB}/\delta|\ll 1$, the LHC dimer is weakly coupled. That is usually the case for two  chlorophylls, $A$ and $B$, in different bio-systems. For a weakly coupled LHC dimer, in the NPQ mode, the $ChlB^*$ may play a role of rapid re-population of the exciton inside the LHC dimer, due to the strong  (``resonant") interaction with the environment.  When $|V_{AB}/\delta|\gtrsim 1$, the dimer is most pronounced (strongly coupled) and has two (symmetric and antisymmetric) states. In this case, the interference effects can increase the efficiency of the NPQ performance. But additional experiments are needed to verify that a strongly couple LHC dimer is indeed realized in the NPQ mode.

We also mention that, as our numerical results demonstrate, in the NPQ mode quantum coherent effects occur. This can be studied by using recently available 2D femtoseconds spectroscopy, and other techniques.

Different generalizations of our model are possible. Namely, our model can be easily generalized for more complicated networks of light-sensitive molecules, heterodimers, charge states, and damaging channels. The protein environment can be modeled by using the hybrid approaches, which are now available. Because in the NPQ mode, not only single but multi-exciton states could play a role, they also should be included in the future research. The future research should also include a detailed analysis of the structures of sinks. Namely, instead of characterizing the sinks by such phenomenological parameters as rates and efficiencies, the studies of the interfaces between the LHC and sinks and the internal structures of the sinks will provide better understanding for the NPQ processes. This research is now in progress.

We believe that our results will be useful for better understanding the NPQ processes in natural bio-complexes and for engineering  artificial NPQ systems.

\ack

This work was carried out under the  auspices of the National Nuclear Security Administration of the U.S. Department of Energy at Los Alamos National Laboratory under Contract No. DE-AC52-06NA25396. We thank Gary Doolen for useful remarks. A.I.N. acknowledges the support from the CONACyT, Grant No. 15349, and a partial support during his visit to the CNLS and the Biology Division, B-11, at LANL. S.G. acknowledges the support from the Israel Science Foundation under grant No. 711091. G.P.B. and R.T.S. acknowledge support from the LDRD program at LANL.

\appendix
\section{}

\subsection*{The non-Hermitian Effective Hamiltonian}

We consider the time-dependent Hamiltonian of the $N$-level system coupled with independent sinks (with continuum spectra) through each level:
\begin{eqnarray}
\mathcal H = &\sum^N_{n=1} E_n|n\rangle\langle n| + \sum_{m, n} \beta_{nm}(t)|n\rangle\langle m|  \nonumber \\
&+  \sum^{N}_{n=1}\sum^{N_n}_{i_n=1} ( E_{i_n}|i_n\rangle\langle  i_n | +V_{ni_n}|n\rangle\langle  i_n | + V_{i_n n}|i_n\rangle\langle n |\big),
\end{eqnarray}
where $m,n = 1,2, \dots,N$. We assume that the sinks are sufficiently dense, so one can perform an integration instead of a summation over sink states. We have,
\begin{eqnarray}\label{ah1}
\mathcal H = &\sum_{n}E_n(|n\rangle\langle n| + \sum_{m,n} \beta_{mn}(t)|n\rangle\langle m| 
+\sum_{n}\int E|E\rangle\langle E|g_n(E)dE  \nonumber \\
&+\sum_{n} \Big( \int\alpha_n(E)|n\rangle\langle E|g_n(E) dE + \rm h.c.\Big),
 \end{eqnarray}
 where $g_n(E)$ is the density of states, and $V_{ni_n}\rightarrow \alpha_n(E)$.

With the wave function written as,
\begin{eqnarray}\label{aS1}
|\psi(t)\rangle = \sum_{n}\Big(c_n(t) |n\rangle + \int c_{nE}(t)|E\rangle g_n(E)dE \Big ),
\end{eqnarray}
the Schr\"odinger equation,
\begin{eqnarray}\label{ach10}
    i\frac{\partial |\psi(t) \rangle}{\partial t} = {\mathcal H(t)} |\psi(t) \rangle,
\end{eqnarray}
takes the form,
\begin{eqnarray}\label{ach11}
    i\dot c_n(t)=& E_n c_n(t) + \sum_{m} \beta_{nm}(t) c_m(t) + \int\alpha_n^\ast(E) c_{nE}(t)g_n(E) dE, \\
    i\dot c_{nE}(t)=&E c_E(t) + \alpha_n(E) c_n(t) .
    \label{ach11a}
\end{eqnarray}
Equation (\ref{ach11a}) can be formally integrated with respect to time,  taking $c_{nE}(0) =0$. Substituting the result in Eq. (\ref{ach11}), we obtain the following system of integro-differential equations, describing the non-Markovian dynamics of the quantum system:
\begin{eqnarray}\label{ach17a}
\fl
    i\dot c_n(t)= E_n c_n(t) + \sum_{m} \beta_{nm}(t) c_m(t)
     -i \int^{t}_{0}  c_{n}(s) e^{-iE(t-s)} ds\int {|\alpha_n(E)|^2g_n(E) \, dE }.
\end{eqnarray}
Transforming to the interaction representation with $c_n(t) = a_n(t)e^{-iE_n t- i\varphi_n (t)}$, where $\varphi_n (t)= \int_0^t \beta_{nn}(\tau) d\tau$, we obtain
\begin{eqnarray}\label{A6}
    i\dot a_n(t)=  \sum_{ m\neq n} \beta_{nm}(t) a_m(t) e^{-i\Delta_{mn} (t)} \nonumber \\
     -i \int^{t}_{0}  a_{n}(s) e^{-i(E-E_n)(t-s)} e^{- i(\varphi_n (t) - \varphi_n (s))} ds\int {|\alpha_n(E)|^2g_n(E) \, dE },
\end{eqnarray}
where $\Delta_{mn}(t) = (E_m -E_n) t +\varphi_m (t) - \varphi_n (t)$.

It is not possible to proceed analytically further without additional assumptions. In what follows, we use the Markov approximation, which can be justified under the following assumptions. Let us assume that, $g_n(E)$ and $|\alpha_n(E)|$, are  slowly varying functions of $E$. Then, the main contribution to the integral, $\int (\dots)dE$, occurs in  the time-interval $|t-s| \lesssim 2\pi/\delta_n$, where, $\delta_n$, is bandwidth of the band with the density of states $g_n(E)$.  In the Markov approximation, only a contribution at $t = s$ in the integral on the r.h.s. of Eq. (\ref{A6})  can be taken into account. This corresponds to the limit,  $\delta_n \rightarrow \infty$. (See discussion in Ref. \cite{KND}.) 

Using the Markov approximation, we obtain, 
\begin{eqnarray}\label{A7}
\fl
    i\dot a_n(t)=  \sum_{ m \neq n} \beta_{nm}(t) a_m(t) e^{-i\Delta_{mn} (t)}
     -i a_{n}(t)\int^{t}_{0}   e^{-i(E-E_n)(t-s)} ds\int {|\alpha_n(E)|^2g_n(E) \, dE },
\end{eqnarray}

Since for $|t-s| > 2\pi/\delta_n$, the contribution in the integral over $E$ is zero, one can approximate the integral over $s$ as follows \cite{KND}:
\begin{eqnarray}\label{A8}
\fl
\int^{t}_{0}   e^{-i(E-E_n)(t-s)} ds\approx  \int^{\infty}_{0}   e^{-i(E-E_n)\tau} d\tau =
-i {\mathcal P}  \bigg\{ \frac{1}{E-E_n }\bigg\} +\pi\delta(E-E_n),
\end{eqnarray}
where ${\mathcal P}$ denotes the  principal value. Substituting this result in (\ref{A7}), we find,
\begin{eqnarray}\label{A9}
 i\dot a_n(t)=  \sum_{m \neq n} \beta_{nm}(t) a_m(t) e^{-i\Delta_{mn} t}
     - a_{n}(t)(\Delta(E_n) +\frac{i}{2}\Gamma_n,
\end{eqnarray}
where
\begin{eqnarray}\label{A10}
\Delta(E_n)&=\mathcal {P} \int \frac{|\alpha_n(E)|^2 g_n(E) dE }{E - \epsilon_n}, \\
 \Gamma_n& =
  2\pi\int{|\alpha_n(E)|^2 }g_n(E){\delta(E -E_n)}\, dE
  = 2\pi g_n(E_n)|\alpha_n(E_n)|^2 .
\end{eqnarray}

Returning to the variables, $c_n(t)$, we obtain,
\begin{eqnarray}\label{A11}
    i\dot c_n(t)=  \varepsilon_n c_n(t) + \sum_{m}  \beta_{nm}(t) c_m(t)  - \frac{i\Gamma_n}{2}  c_n(t),
\end{eqnarray}
where $\Gamma_n = \Gamma_n(E_n)$ and $\varepsilon_n = E_n  -\Delta(E_n)$.

Representing the wave function, $|\psi_N(t)\rangle = \sum_{n}c_n(t) |n\rangle $, we find that the dynamics of the $N$-level system interacting with the continuum spectra ($N$ independent sinks) is described by the Schr\"odinger equation,
\begin{eqnarray}\label{ach10a}
    i\frac{\partial |\psi_N(t) \rangle}{\partial t} =\tilde {\mathcal H} |\psi_N(t) \rangle,
\end{eqnarray}
where, $ \tilde{\mathcal H}= {\mathcal H}- i \mathcal W$, is the effective non-Hermitian Hamiltonian, with the Hermitian part represented by the dressed Hamiltonian,
 \begin{eqnarray}
 {\mathcal H} = \sum_{n}\varepsilon_n|n\rangle \langle n|  +\sum_{m, n}  \beta_{mn}(t)|m\rangle \langle n|
 \end{eqnarray}
and the non-Hermitian part is,
 \begin{eqnarray}
  W = \frac{1}{2}\sum_{n}\Gamma_n|n\rangle \langle n|.
 \end{eqnarray}

 Equivalently, the dynamics of this system can be described by the generalized Liouville equation,
 \begin{eqnarray}\label{DM1}
    i \dot{ \rho} =\tilde {\mathcal H}\rho - \rho{\tilde {\mathcal H}}^\dagger= [\mathcal H,\rho] - i\{\mathcal W,\rho\},
 \end{eqnarray}
 where   $\rho $ is the density matrix projected on the discrete states, and $\{\mathcal W,\rho\}= \mathcal W\rho +\rho\mathcal  W$.

\section{}
\subsection*{Equations of motion in the presence of noise}

We consider the $N$-level system with each level coupled to an independent sink (modeled by a continuum energy spectrum).  In the Weisskopf-Wigner pole approximation \cite{SM,WW,scully}, the $N$-level system interacting with the continuum is governed by the effective non-Hermitian Hamiltonian,  $ \tilde{\mathcal H}= {\mathcal H}- i \mathcal W$, where,
 \begin{eqnarray}
{\mathcal W} = \frac{1}{2}\sum_{n}\Gamma_n|n\rangle \langle n|,
 \end{eqnarray}
 and the $\Gamma_n$, are the tunneling rates to the sinks. (For details see Appendix A.)

The dynamics of this system can be described by the generalized Liouville  equation,
 \begin{eqnarray}\label{A2}
    \dot{ \rho} = i[\rho,\mathcal H] - \{\mathcal W,\rho\},
 \end{eqnarray}
 where $\{\mathcal W,\rho\}= \mathcal W\rho +\rho\mathcal  W$.
To include the effects of a thermal bath, we use the reduced density matrix approach leading to
the master equation:
\begin{eqnarray}\label{Meq1}
\frac{d\rho}{dt} =i[{\rho,\mathcal H}] - \{\mathcal W,\rho\}+{\mathcal L}\rho,
\end{eqnarray}
where  the superoperator, ${\mathcal L}$, describes a coupling to the bath.

We  assume that the Hamiltonian, ${\mathcal H(t)}$, depends on some control
parameters, $\lambda_i$. The noise associated with  fluctuations of these parameters is described by
random functions,  $\delta\lambda_i(t)$. For simplicity, we restrict ourselves to only one
fluctuating parameter, $\delta\lambda(t)$, denoting it as, $\xi(t)$. The generalization to many
parameters is straightforward. Expanding  the Hamiltonian to first order in $\xi(t)$, we obtain,
\begin{eqnarray} \label{Neq1}
{\mathcal H} = {\mathcal H}_0 +{\mathcal V}\xi(t).
\end{eqnarray}

Using (\ref{Neq1}), one can represent Eq. (\ref{Meq1}) as,
\begin{eqnarray}\label{Meq2a}
\frac{d\rho(t)}{dt} =i[\rho(t),{\mathcal H}_0 ] +
i [\rho(t),\xi(t)  {\mathcal V}]-  \{\mathcal W,\rho\} + {\mathcal L}\rho(t) .
\end{eqnarray}
For the average density matrix, this yields,
\begin{eqnarray}\label{A3a}
\frac{d\langle\rho(t)\rangle}{dt} =i [\langle\rho(t)\rangle,{\mathcal H}_0] 
 +i[\langle  X(t)\rangle,{\mathcal V} ] -  \{\mathcal W,\langle\rho(t)\rangle\}+{\mathcal L}\langle\rho(t)\rangle  ,
 \end{eqnarray}
where $\langle  X(t)\rangle= \langle { \xi}(t) \rho(t)\rangle $, and the average $\langle
{\dots}\rangle$ is taken over the random process describing the noise.

Further we assume that fluctuations are produced by the random telegraph process (RTP) with the correlation function,  $\chi(\tau) =\sigma^2e^{-2\gamma \tau}$. Employing the differential formula for RTP \cite{KV1,KV2,KV3},
\begin{eqnarray}
\Big(\frac{d}{dt} +2\gamma \Big)\langle  {\xi}(t)R[t;\zeta(\tau) ]
\rangle =\Big\langle  {\xi}(t)\frac{d}{dt}R[t;\xi(\tau)]
\Big\rangle ,
\end{eqnarray}
where, $R[t;\xi(\tau)]$, is an arbitrary functional, we obtain from Eq. (\ref{A3a}) the following closed system of differential equations:
\begin{eqnarray}\label{A4a}
\fl
\frac{d\langle\rho(t)\rangle}{dt} =&i [\langle\rho(t)\rangle,{\mathcal H}_0] 
 +i[\langle  \rho^\xi (t)\rangle,\Lambda  ] -  \{\mathcal W,\langle\rho(t)\rangle\}+{\mathcal L}\langle\rho(t)\rangle  , \\
 \fl
 \frac{d\langle\rho^\xi(t)\rangle}{dt} =&i [\langle\rho^\xi(t)\rangle,{\mathcal H}_0] 
 +i[\langle  \rho (t)\rangle,\Lambda  ] -  \{\mathcal W,\langle\rho^\xi(t)\rangle\} - 2\gamma \langle\rho^\xi\rangle+{\mathcal L}\langle\rho^\xi(t)\rangle ,
 \label{A4b}
\end{eqnarray}
where $\langle  \rho^\xi (t)\rangle= \langle \xi (t)  \rho (t)\rangle/\sigma$ and $\Lambda   = \sigma {\mathcal V} $.

\section*{References}
\addcontentsline{toc}{section}{References}


\end{document}